\documentclass[12pt,preprint,revtex4]{aastex}
\usepackage{epstopdf,graphicx,mathtools,amsmath,amssymb}

\newcommand\WCO{\ifmmode{W_\mathrm{CO}\,}\else{$W_\mathrm{CO}$\,}\fi}
\newcommand\htwo{\ifmmode{{\rm H}_2}\else{H$_2\,$}\fi}
\newcommand{\hi}{H\,{\small I}\,}
\newcommand{\hii}{H\,{\small II}\,}
\newcommand\xco{\ifmmode{X'_\mathrm{CO}\,}\else{$X'_\mathrm{CO}$\,}\fi}
\newcommand\Av{\ifmmode{A_V}\else{$A_V$}\fi}
\newcommand\pers{\ifmmode{\rm s^{-1}}\else{s$^{-1}$}\fi}
\newcommand\km{\ifmmode{\rm km}\else{km}\fi}
\newcommand\kms{\km\,\pers}
\newcommand\cmtwo{\ifmmode{\rm cm^{-2}}\else{cm$^{-2}$}\fi}
\newcommand\EBV{\ifmmode{E(B-V)}\else{$E(B-V)$}\fi}
\newcommand\EBVres{\ifmmode{E(B-V)_\mathrm{res}\,}\else{$E(B-V)_\mathrm{res}$\,}\fi}
\newcommand\Avres{\ifmmode{A_{V, \mathrm{res}}\,}\else{$A_{V,\mathrm{res}}$\,}\fi}
\newcommand\jone{\ifmmode{J\!=\!1\!\rightarrow\!0}\else{$J\!=\!1\!\rightarrow\!0$}\fi}
\newcommand{\ee}[1]{\ifmmode{\times 10^{#1}}\else{$\times 10^{#1}\,$}\fi}
\newcommand{\xav}{\ifmmode{X'_\mathrm{A_V}}\else{$X'_\mathrm{A_V}$}\fi}

\begin{document}
\title{High Latitude, Translucent Molecular Clouds as Probes of Local Cosmic Rays}

\author{Ryan D. Abrahams$^{1,2,3}$ and Timothy A. D. Paglione$^{1,2,3}$}
\email{rabrahams@gc.cuny.edu}
\altaffiltext{1}{Department of Physics, Graduate Center of the City
University of New York, 365 Fifth Ave., New York, NY 10016, USA}
\altaffiltext{2}{Department of Earth \& Physical Sciences, York
College, City University of New York, 94-20 Guy R. Brewer Blvd.,
Jamaica, NY 11451, USA}
\altaffiltext{3}{Department of Astrophysics, American Museum of
Natural History, Central Park West at 79th Street, New York, NY 10024, USA}

\begin{abstract}

We analyze the gamma-ray emission from 9 high latitude, translucent molecular clouds taken with the \textit{Fermi} Large Area Telescope (LAT) between 250 MeV and 10 GeV. Observations of gamma-rays allow us to probe the density and spectrum of cosmic rays in the solar neighborhood. The clouds studied lie within $\sim\!270$ pc from the Sun and are selected from the \textit{Planck} all-sky CO map. Gamma-rays in this energy range mostly result from cosmic ray interactions with the interstellar medium, which is traced with three components: \hi, CO, and dark gas. Every cloud is detected and shows significant, extended gamma-ray emission from molecular gas. The gamma-ray emission is dominated by the CO-emitting gas in some clouds, but by the CO-dark gas in others. The average emissivity and gamma-ray power law index from \hi\ above 1 GeV shows no evidence of a systematic variation. The CO-to-\htwo\ conversion factor shows no variation between clouds over this small spatial range, but shows significant variations within each cloud. The average CO-to-\htwo\ conversion factor suggests that the CO-dark gas is molecular as opposed to optically thick \hi.

\end{abstract}


\section{Introduction}
Molecular clouds at high galactic latitude represent an abundant source of gamma-rays due to interactions with cosmic rays \citep{Digel1996, Abdo2010,  Ackermann2012Cyg, Ackermann2012c, AckermannOrion, Marti2013}. High latitude clouds are primarily low mass and harbor little to no active star formation (see \citealt{McGehee2008} for a review). They are also often relatively isolated from localized cosmic ray acceleration sites such as supernova remnants or OB associations. Thus, their gamma-ray emission should reflect the steady-state cosmic ray density and spectrum of the surrounding region in the Galaxy. These clouds are all nearby with most having a distance $d \la 350$ pc for $|b|> 25^\circ$ given a scale height of 150 pc \citep{Magnani1996}. High latitude clouds therefore represent potentially pristine probes of the cosmic ray spectrum in the solar neighborhood. 

Diffuse gamma-ray emission comes from a combination of cosmic ray interactions with the interstellar medium (ISM), inverse Compton scattering of ambient radiation by cosmic ray electrons and positrons, and extragalactic diffuse emission. Cosmic ray interactions with the ISM produce primarily GeV gamma-rays via proton-proton collisions which lead to the production of neutral and charged pions. Neutral pions decay directly into gamma-rays while the charged pion species ultimately decay into electrons and positrons. Cosmic ray leptons interact with gas to create gamma-rays mainly via bremsstrahlung emission. The ISM is effectively transparent to cosmic rays, so the gamma-ray emission is sensitive to the total gas column density regardless of dust properties or gas state (\hi, \hii, \htwo). 

The necessary likelihood modeling of \textit{Fermi} Large Area Telescope (LAT) data incorporates contributions to the gamma-ray emission from gas traced by 21 cm \hi\ emission and the integrated 2.6 mm CO(\jone) line emission (\WCO) \citep{Lebrun1983}. \WCO\ is assumed to be directly proportional to the column density of molecular hydrogen: $N(\htwo) = X_\mathrm{CO} \WCO$. In fact, gamma-rays have been used to calibrate $X_\mathrm{CO}$ \citep[e.g.,][]{Bloemen1984}. However, likelihood models of gamma-ray emission based solely on the distribution of these two species exhibit significant residual gamma-ray emission \citep{Grenier2005, Ackermann2011}. This excess gamma-ray emission is assumed to trace unseen molecular or atomic gas, which has been called dark gas. In cold molecular clouds, the dark gas mass can be a significant fraction of the total gas mass\citep{Ade2011,Paradis2012,Pineda2013}. 

Dark gas is expected in photodissociation regions and diffuse clouds exposed to the interstellar radiation field. Because CO self-shields less efficiently and has a lower dissociation energy than \htwo, CO dissociation occurs deeper into a cloud (i.e., to higher \Av). Therefore, between $1 <\Av < 5$, CO fails to trace \htwo\ linearly \citep{vanDishoeck1988, Wolfire2010}. Clouds in this range of \Av\ are classified as ``translucent" \citep{vanDishoeck1988}, and represent most molecular clouds at high latitude \citep{Magnani1996}. Translucent clouds tend to be smaller and less dense than giant molecular clouds, but should be much more numerous \citep{MBM85}.

A lingering problem with modeling gamma-rays from high latitude clouds had been that existing CO maps were unavailable, sparse, or incomplete. The \textit{Planck} CO map is the first all-sky map of the CO( \jone\ ) emission line \citep{Ade2013CO}. We use this CO map in a complete, flux-limited survey of high latitude molecular clouds with the \textit{Fermi} LAT. In this paper, we describe the methods and systematic uncertainties of the survey and report on gamma-ray observations of 9 high latitude, translucent molecular clouds. Every cloud has detectable gamma-ray emission consistent with maps of either CO, dark gas, or both. We also report the photon index of each cloud, as it reflects the incident cosmic ray spectrum. This work presents the initial results from the full survey.

\section{Source Selection}

For this pilot survey, we identify well-known molecular clouds  with $|b| > 25\degr$ from previous surveys to analyze in gamma-rays. Clouds at such latitudes are all nearby, which ensures the gamma-ray flux is high enough to study. Large clouds at lower galactic latitude, such as Orion, have already been extensively studied with the \textit{Fermi} LAT \citep{Ackermann2012c, Ackermann2012a}.

\citet{Torres2005} presented a study on the possible gamma-ray emission based on a number of CO surveys \citep{Magnani1996, Hartmann1998, Magnani2000, Dame2001}. We find candidate clouds from these surveys and identify them in the \textit{Planck} CO map. We choose several of the brightest high latitude clouds from these surveys. Three other, fainter clouds are chosen to explore the low \WCO limits for gamma-ray detection. Two bright clouds, G313.1-28.6 and G315.1-29.0, are identified via visual inspection of the \textit{Planck} CO map. These were observed in an earlier CO catalog towards dark clouds \citep{Otrupcek2000} and named Chamaeleon-East {\small II} \citep{Mizuno2001}, but were not part of the surveys described in \citet{Torres2005}, nor mapped by \citet{Dame2001}.

In addition, we choose a region devoid of large-scale gas and dust emission by visual inspection of the \textit{Planck} CO map and the color excess map of \citet{Abergel2014}. This region, centered on Galactic coordinates $(\ell,b)=(250^\circ,30^\circ)$, is used to test the false detection rate of gamma-ray emission from CO or dark gas.

Some relevant properties of the chosen clouds are listed in Table \ref{tab:prop}. The areas of the clouds are calculated from the CO extent in the \textit{Planck} map. Distances to the clouds are taken from the literature and were derived from color excess \citep{Schlafly2014, Lallement2013}. Many, but not all, of the clouds have masses estimated in prior surveys, \citep[e.g.,][]{Magnani1996}.

\begin{figure}[h]
	\centering
	\includegraphics[width=\linewidth]{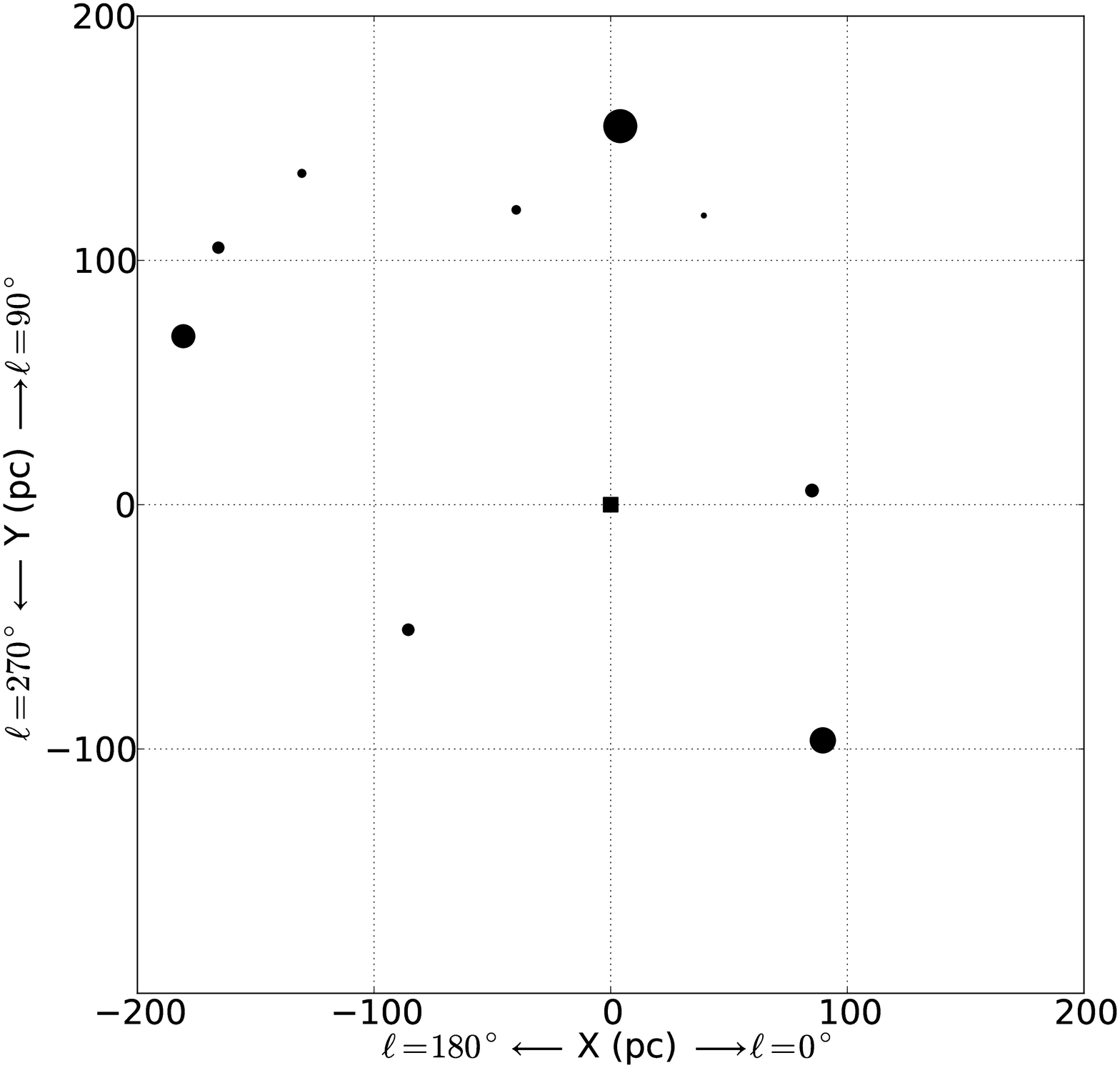}
	\caption{Distances of the clouds from the Solar System, marked with a square in the middle, projected onto the galactic plane. The size of the dot corresponds to the angular size of the cloud.}
	\label{fig:dist_xy}
\end{figure}

\begin{figure}[t]
	\centering
	\includegraphics[width=\linewidth]{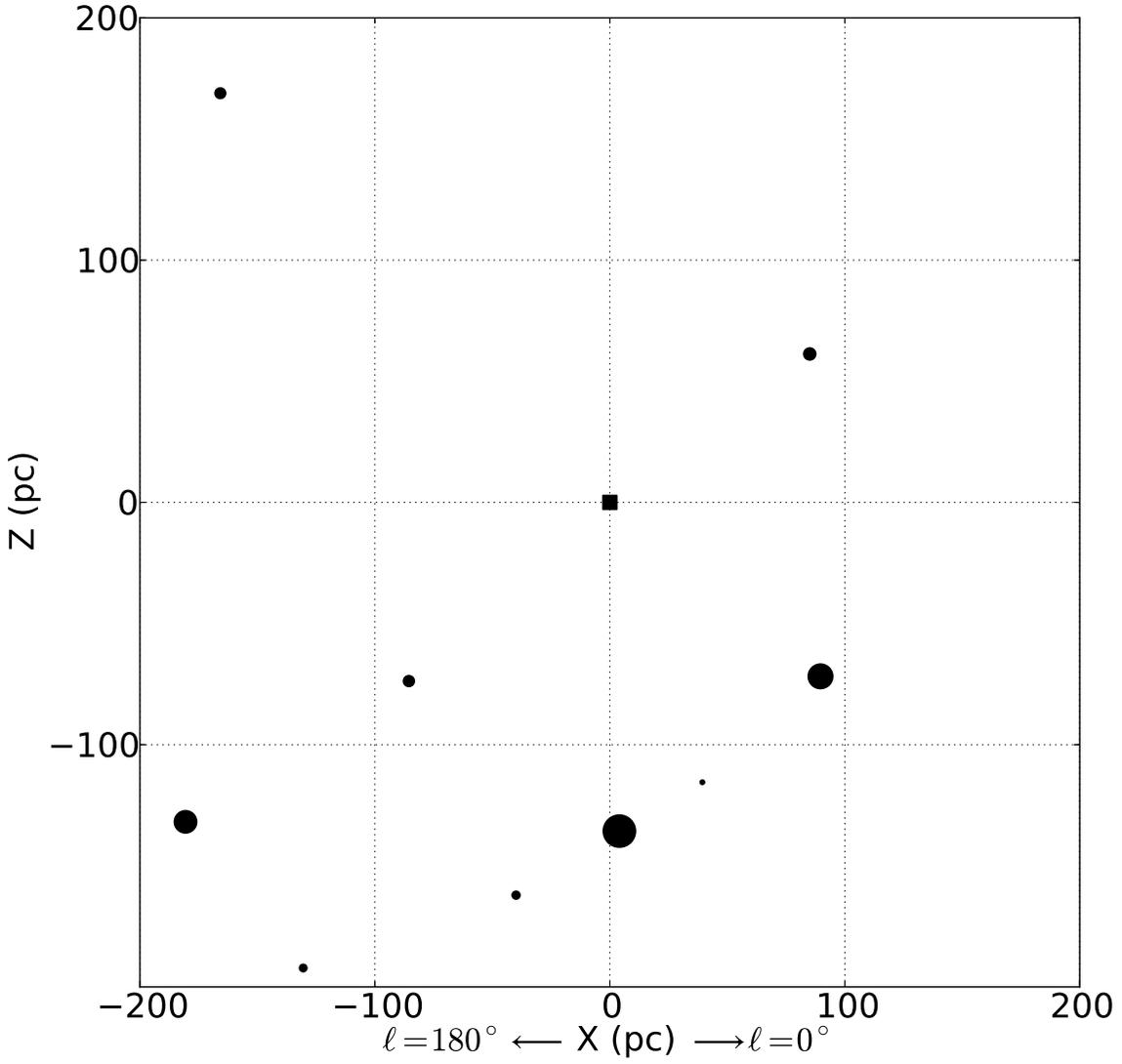}
	\caption{Same as Figure \ref{fig:dist_xy}, except the vertical direction indicates distance from the galactic plane and the Galactic Center lies off to the right.}
	\label{fig:dist_xz}
\end{figure}

\begin{figure}[t]
	\centering
	\includegraphics[width=\linewidth]{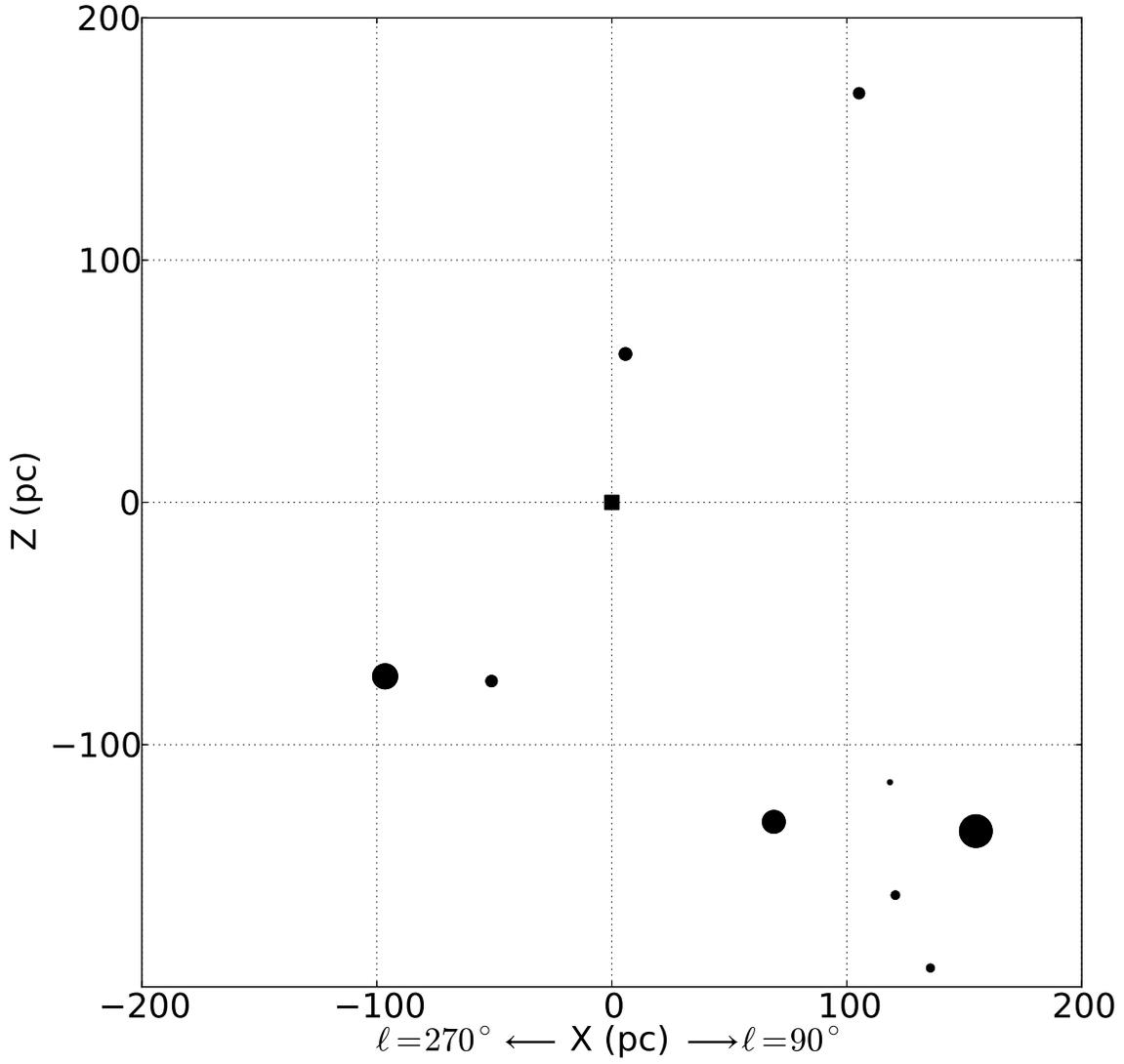}
	\caption{Same as Figure \ref{fig:dist_xy}, except the vertical direction indicates distance from the galactic plane.}
	\label{fig:dist_yz}
\end{figure}

The locations of the clouds relative to the Solar System can be seen in Figures \ref{fig:dist_xy}, \ref{fig:dist_xz}, and \ref{fig:dist_yz}. Most of the clouds lie below the galactic plane, as shown in Figures \ref{fig:dist_xz} and \ref{fig:dist_yz}, but represent a large range of Galactic longitudes seen in Figure \ref{fig:dist_xy}.

\section{Gamma-ray Analysis}

The LAT on the \textit{Fermi Gamma-ray Space Telescope} is a pair-tracking telescope, sensitive to gamma-rays between 20 MeV and 300 GeV. The tracker is surrounded by anti-coincidence detectors to distinguish between cosmic ray and gamma-ray events. The photon localization strongly depends on the photon energy; at 1 GeV, the 68\% containment radius is 0\fdg8, decreasing with energy to roughly $0\fdg2$ at around 10 GeV \citep{AckermannIRF}.

	We use data of the entire sky from the \textit{Fermi} LAT between August 4, 2008 and June 19, 2013. We use the \textit{Fermi} science tools (v9r27p1) available from the \textit{Fermi} Science Support Center \footnote{FSSC: \url{http://fermi.gsfc.nasa.gov/ssc/}}, utilizing the P7\_V6 instrument response function. When selecting the data, we consider both front and back converted photons in the ``source" class. We select data from a $10\degr$ radius around the chosen coordinates and between 250 MeV and 10 GeV. These energies are chosen to maximize both source localization and photon statistics. Including photons between 10 and 100 GeV does not improve the significance of the detection, as will be explained in Section \ref{sec:results}. We exclude photons with incidence angle $> 100\degr$ from the zenith and any time the spacecraft rocking angle exceeds 52\degr. These constraints remove most gamma-ray contamination coming from the Earth's limb. 

We perform a binned likehihood analysis, selecting the data which lie inside a $14^\circ\times 14^\circ$ square centered on the region of interest (ROI) center. The basic procedure for the likelihood analysis of gamma-ray data is described in \citet{Mattox1996}. To evaluate source detection and model significance, we consider the test statistic ($TS$), which is proportional to the difference of the log likelihoods of two different models:

\begin{equation}
	TS = -2\Big( \ln\mathcal{L}_A - \ln\mathcal{L}_B\Big),
\end{equation}

\noindent where $\mathcal{L}_A$ and $\mathcal{L}_B$ are the likelihoods for two models we are comparing. The $TS$ represents the significance of model $B$ over model $A$. 

To create the model, point sources are taken from the Fermi 2 year catalog \citep{Fermicat2012} version 6 (2FGL). For those clouds with point sources from the 2FGL coincident with the CO emission, we remove the point source from the model. Sources outside of 7\degr\ from the ROI center have all free parameters fixed for the fitting procedure, while closer sources only have their spectral indices fixed. After an initial fitting, weak point sources ($TS < 50$) are removed. New point sources are identified by subtracting the best fit model from the counts map, then smoothing this gamma-ray residual map and identifying regions exceeding 3 standard deviations above the average residual.

Two additional components common to any model include the isotropic and Galactic inverse Compton emission. The isotropic emission, originating from extragalactic diffuse gamma-ray emission and misclassification of cosmic rays in the LAT, is modeled by the ``iso\_p7v6source.txt" provided by the Fermi Science Support Center. The inverse Compton component uses GALPROP \footnote{Sourse code can be found at \url{https://sourceforge.net/projects/galprop}} \citep{GALPROP98,Vladimirov2011}, and is added in as a data cube from the GALPROP input galdef file ``54\_77Xvarh7S" \citep{Ackermann2012c}.

Finally, we model diffuse emission arising from interstellar gas and dust. We use six different models for this analysis to determine the significance of the gamma-ray emission from each gas component. The baseline model, against which we compare all others, contains all the neutral gas elements: \hi, CO, and dark gas (\textit{CODG}). To check the significance of gamma-ray emission from the molecular cloud, we compare the baseline model to one containing only \hi\ (\textit{HI}). Two additional models are used to check the significance of gamma-ray emission from CO and dark gas individually: one model with \hi and dark gas (\textit{DG}) and another model with \hi and CO (\textit{CO}). To check whether the gamma-ray emission comes from an extended source or a point source, the CO and dark gas templates in the model are replaced with a point source located at the peak of the CO emission (\textit{PS}). Finally, we test for any background point sources by adding one at the peak CO emission in addition to \hi, CO, and dark gas (\textit{CODGPS}). In the cases where the ROI has confused 2FGL sources (see table \ref{tab:ptsrces}), we add the specific 2FGL source into the model instead of adding a separate source.

For neutral atomic gas, we separate the \hi\ data from the LAB survey \citep{Kaberla2005} into two templates. In the first, we integrate over the velocity axis between $\pm 20\; \kms$ to account for nearby, ambient gas associated with the cloud. The second template uses the rest of the \hi\ data to account for background \hi\ gas. We calculate the \hi\ column density, $N$(\hi), using a spin temperature $T_S = 125$ K, consistent with previous gamma-ray studies \citep{Abdo2009, Abdo2010, Ackermann2012a,  Ackermann2012c, AckermannOrion}:

\begin{equation}
\label{eq:HIthick}
N({\rm\hi})(v,T_S) = -CT_S \ln\left( 1 - \frac{T_{B}(v)}{T_S - T_{bg}}\right),
\end{equation}

\noindent where $C = 1.83\times 10^{18}$ cm$^{-2}$ K$^{-1}$, and $T_{bg}$ is the background temperature at 1.4 GHz. The $N$(\hi) map is in a local coordinate system centered on the clouds and is linearly interpolated to $0\fdg1$ per pixel from the all sky map.

For the \htwo template, we use \WCO\ maps taken from \textit{Planck} with a $2\sigma$ cut. We use the ``type 2" map, which combines data from different frequency channels to improve overall signal-to-noise by removing contaminating signals from CMB, dust, and free-free emission, though $^{13}$CO contamination remains. The $^{13}$CO contribution is compensated for by dividing \WCO\ by a factor of 1.16 \citep{Ade2013CO, Chen2015}.

In order to make a dark gas template, we remove a linear combination of $N$(\hi) and \WCO\ from the visual extinction (\Av) adapted from the map of color excess, \EBV, from the \citet{Abergel2014} dust model, assuming \Av=$R_V$\EBV, where we use the standard $R_V=3.1$ here \citep{Cardelli1989,Whittet2001}. This residual map, hereafter referred to as \Avres, is estimated as: 

\begin{equation}
	\Avres = \Av - \frac{1}{\alpha}\Big( N({\rm \hi}) + \beta \WCO \Big).
\end{equation}

\begin{figure*}[h]
\begin{minipage}[t]{0.45\textwidth}
	\includegraphics[width=\linewidth]{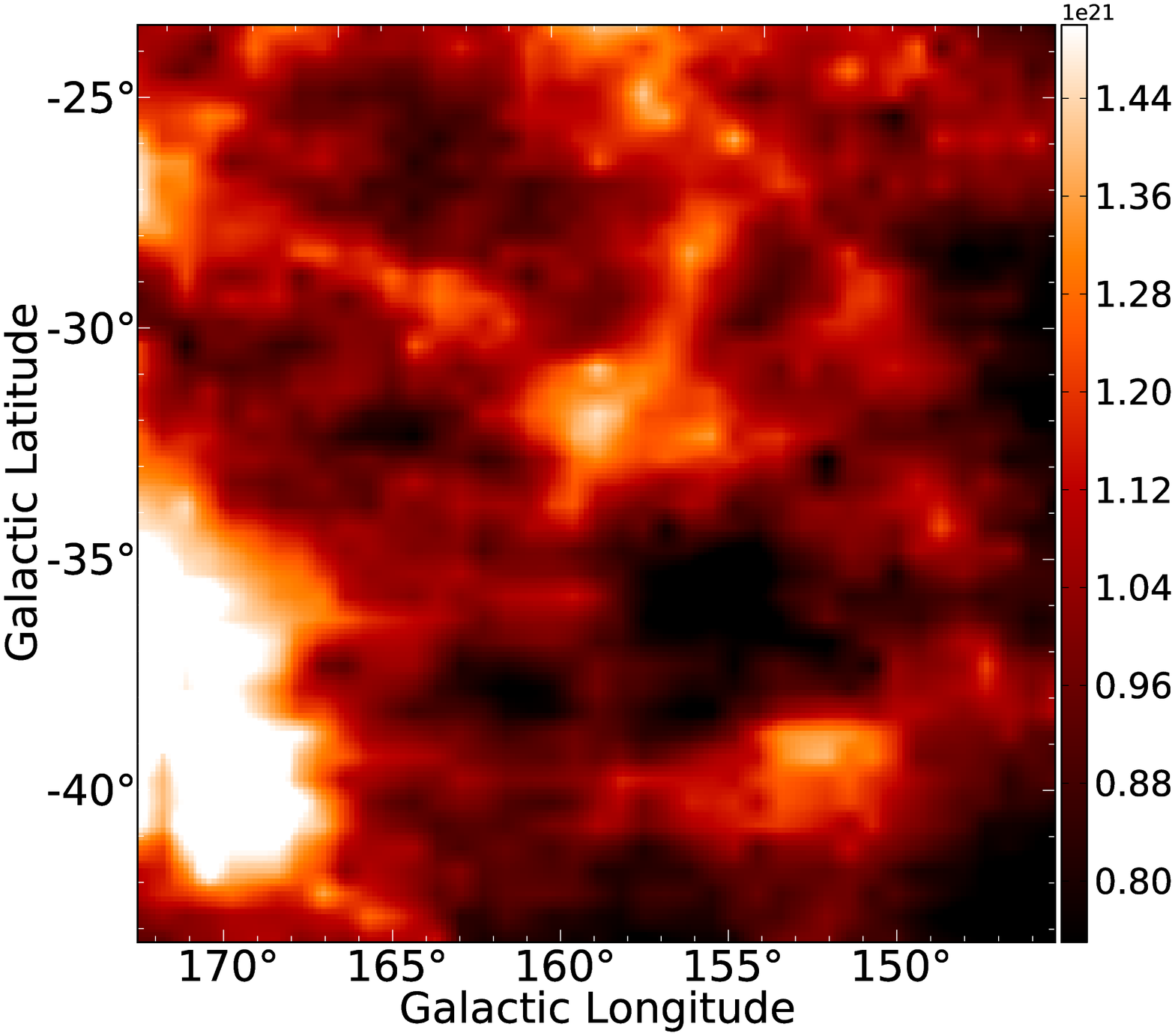}
\end{minipage}
\hspace{\fill}
\begin{minipage}[t]{0.45\textwidth}
	\includegraphics[width=\linewidth]{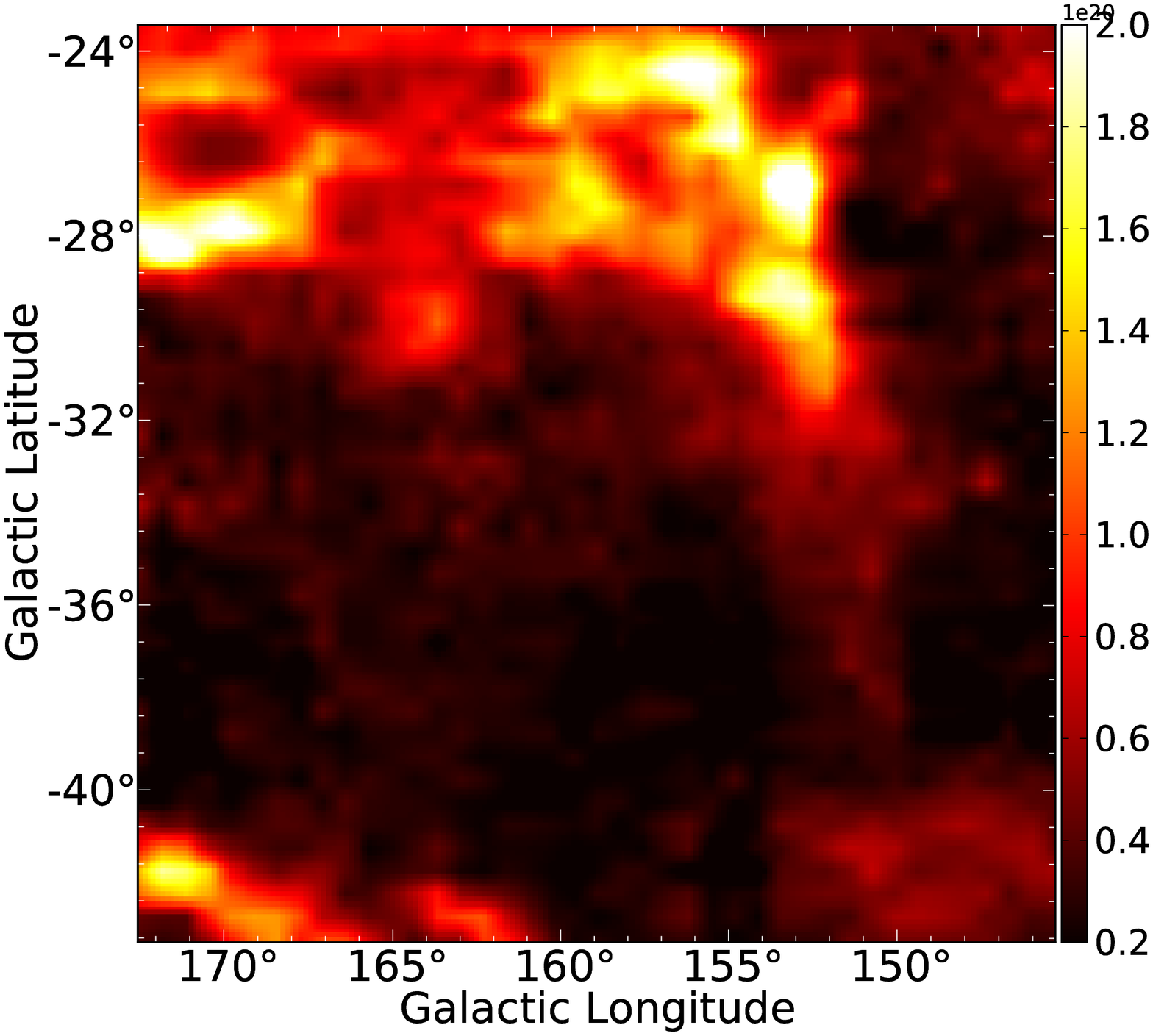}
\end{minipage}

\vspace*{0.5cm}
\begin{minipage}[t]{0.45\textwidth}
	\includegraphics[width=\linewidth]{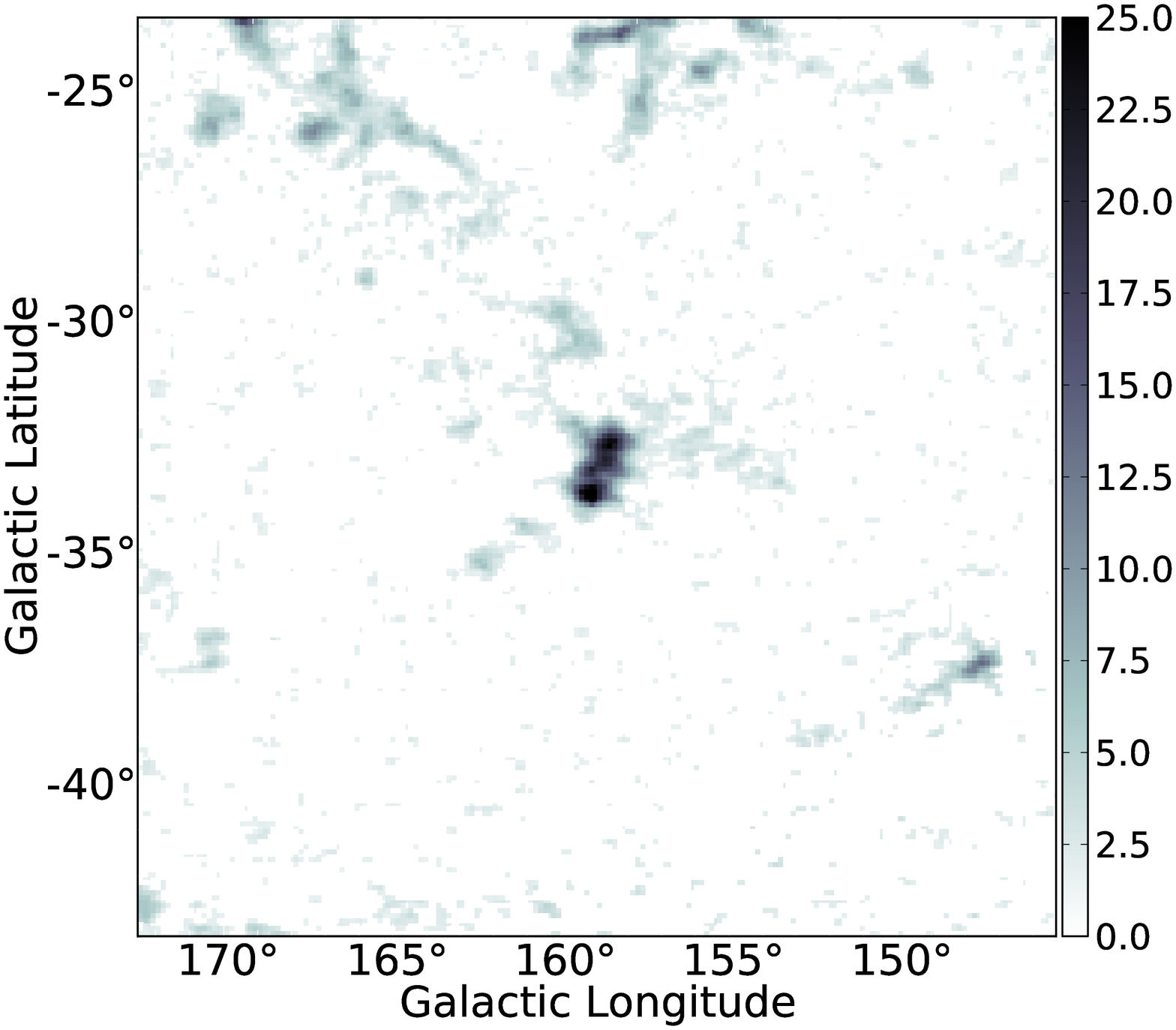}
\end{minipage}
\hspace{\fill}
\begin{minipage}[t]{0.45\textwidth}
	\includegraphics[width=\linewidth]{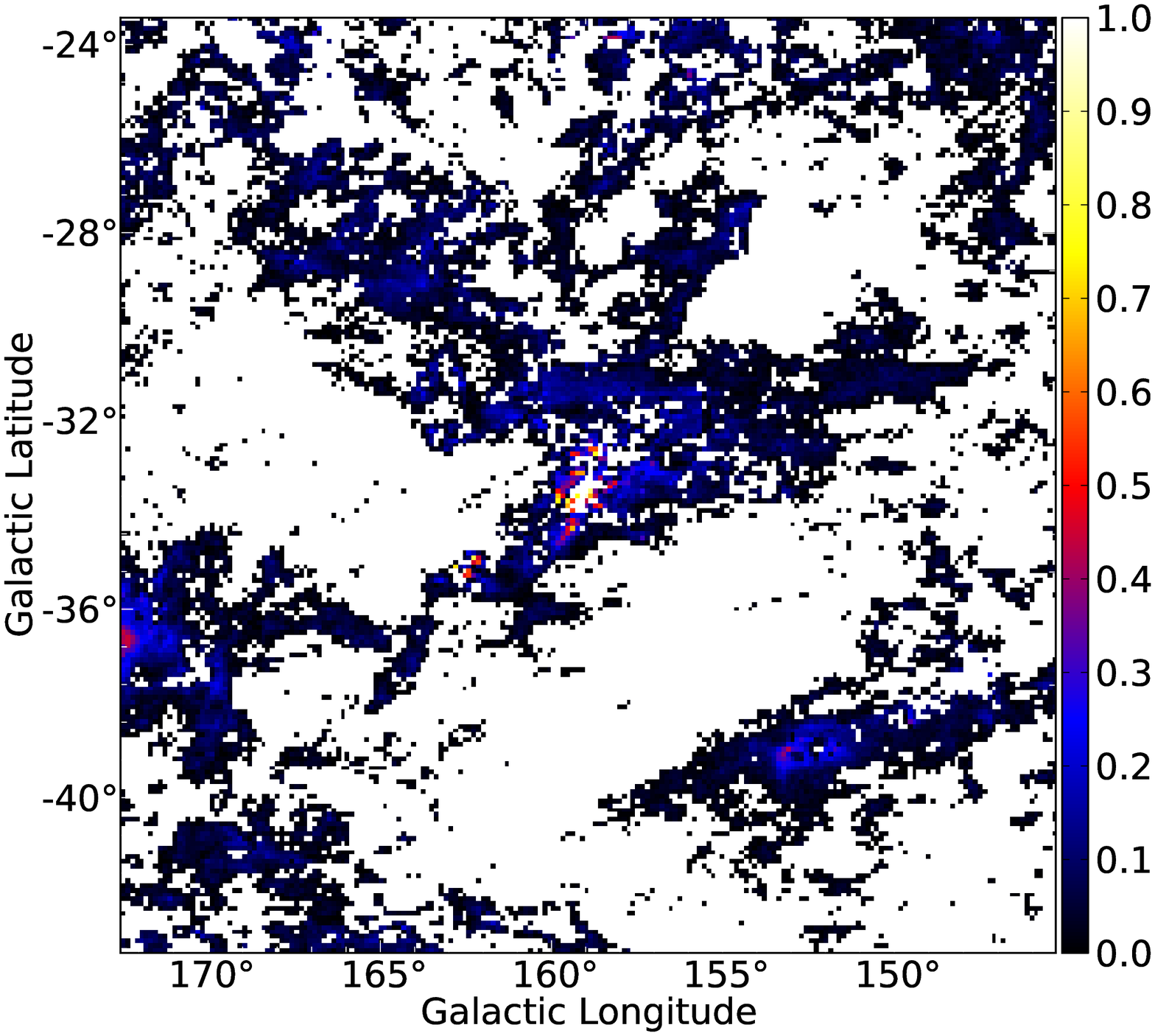}
\end{minipage}
	\caption{Gas templates for MBM 12. \textit{top-left}. Local \hi\ column density in $10^{21}$ cm$^{-2}$, \textit{top-right}. non-local \hi\ column density in $10^{20}$ cm$^{-2}$, \textit{bottom-left}. \WCO in K km s$^{-1}$, \textit{bottom-right}. the dark gas template, \Avres\ in magnitudes.}
	\label{fig:MBM12temp}
\end{figure*}

\noindent The parameters $\alpha$ and $\beta$ are varied until the best fit combination of $N$(\hi) and \WCO\ is found \citep{Grenier2005, Ackermann2012Cyg}. Both the \WCO\ and \EBV\ maps are use the same coordinates as the $N$(\hi) map and are interpolated from the HEALPix maps provided by \textit{Planck}. The four gas templates used for MBM 12 are shown in Figure \ref{fig:MBM12temp}. These templates are all made 3\degr\ larger on each side than the region selected to account for possible gamma-ray photons leaking into the selected region due to the large point-spread function.

Finally, the \textit{Fermi} bubbles \citep{Su2010} represent an additional background source of gamma-ray emission for one cloud in this study, MBM 36. We add a constant component with a power law spectrum across the MBM 36 ROI \citep{Su2010} to account for this emission.

Every model component adds linearly to the total gamma-ray intensity \citep{Ackermann2012Cyg}:

\begin{eqnarray}
	I_\gamma(\ell,b) &= \sum_j PS_j + c_\mathrm{iso} I_\mathrm{iso} + \sum_{i=1}^2 q_{{\rm \hi},i} N_i({\rm \hi})(\ell,b) \nonumber\\ 
	&+ q_\mathrm{CO} \WCO(\ell,b) + q_\mathrm{\Avres}\Avres(\ell,b) + c_\mathrm{IC}I_\mathrm{IC}(\ell,b),
\end{eqnarray}

\noindent where $I_\mathrm{IC}$ is the inverse Compton contribution to the gamma-ray intensity, $I_\mathrm{iso}$ is the isotropic contribution, and $PS$ is the contribution due to point sources. The fit parameters are $c_\mathrm{iso}$ and $c_\mathrm{IC}$, which are normalization coefficients, and $q_\mathrm{\hi}, q_\mathrm{CO}$, and $q_\mathrm{\Avres}$ which are the emissivities of the respective gas templates in units of photon flux per H-atom \cmtwo, per K \kms, and per magnitude of residual extinction, respectively.

To perform the fitting, we used the Python likelihood analysis tools provided by the FSSC. Using gtlike, we find an approximate solution with the `DRMNFB' optimizer, and then refine the solution by fixing point sources with $TS<100$, removing any source with $TS<50$, and refitting with the `NewMinuit' optimizer.

\section{Results}\label{sec:results}

We find diffuse gamma-ray emission from molecular gas to high significance in all 9 regions studied. The control region has no diffuse gamma-ray emission from CO or dark gas. All clouds show extended gamma-ray emission, and all four 2FGL associations listed in Table \ref{tab:ptsrces} are identified as diffuse gamma-ray emission from the clouds themselves. Only MBM 12 shows evidence of a background AGN. MBM 12, given its provisional classification as a blazar in the 2FGL and potential association with  a radio source, goes through additional analyses discussed in a later subsection.

Table \ref{tab:TSvals} lists the $TS$ values for each model compared to the baseline model \textit{CODG}. Each model was fit separately to the data. The significance of gamma-ray emission from both CO and dark gas is given as $TS_\mathrm{\htwo}$ by comparing model \textit{CODG} to \textit{HI}. The significance of gamma-ray emission from the dark gas template is given as $TS_\mathrm{DG}$ by comparing model \textit{CODG} to \textit{CO}. The significance of gamma-rays from CO-emitting gas is given as $TS_\mathrm{CO}$ by comparing model \textit{CODG} to model \textit{DG}. Determining whether any gamma-ray emission comes from an extended source is given as $TS_\mathrm{ex}$ by comparing model \textit{CODG} to model \textit{PS}. Finally, we test for contributions due to a background point source, presumably an AGN, given as $TS_\mathrm{AGN}$, by comparing model \textit{CODG} with model \textit{CODGPS}. A $TS>20$ indicates a significant difference between the two models tested.

In the control region, $TS_\mathrm{H_2}= 1$, indicating that we do not detect gamma-rays from gas traced by CO or dark gas. For the ROIs around the clouds MBM 04 and DIR 071-43 there is no evidence for significant gamma-ray emission from gas traced by CO, as shown by a low $TS_{\rm CO}$. These ROIs show evidence for gamma-ray emission dominated by dark gas. Conversely, the ROIs for MBM 55 and MBM 32 show significant gamma-ray emission from gas traced by CO but not by dark gas; the $TS_{\rm CO}$ is high while the $TS_{\rm DG}$ is low. The remainings ROIs show evidence for emission from both gas traced by CO and dark gas to different degrees. For example, the ROI around MBM 12 show significantly more gamma-ray emission from gas traced by CO as opposed to dark gas, while the ROI around MBM 36 shows more gamma-ray emission from dark gas. Since this sample contains clouds dominated by gamma-ray emission from CO and some dominated by emission from dark gas, the translucent clouds may represent the transition to fully molecular clouds.

For every ROI, $TS_\mathrm{ex}$ is large enough to conclude that the gamma-ray emission is not coming from a single point source. A $TS_\mathrm{AGN}>20$ indicates the presence of a background AGN. Only MBM 12, discussed in the following section, shows significant evidence for a  background source. Thus, out of the four point sources given in Table \ref{tab:ptsrces}, we confirm the existence of one background gamma-ray point source and attribute the other three to diffuse gamma-ray emission from molecular clouds. 

We also compare the baseline model to one utilizing the \textit{Fermi} standard Galactic diffuse model for Pass 7\footnote{gal\_2yearp7v6\_source}. However, only one cloud in our survey had CO data from \citet{Dame2001} included in the diffuse model (MBM 12). Therefore our models, which include CO, should more accurately reflect the gas origin of the gamma-ray emission.  Further, as will be discussed in Section \ref{sec:uncertain}, the dust alone results in a poor fit to the data.

Over the entire energy range, when our gas templates all assume a single power law spectrum, the standard diffuse model has a higher likelihood than any of the models we fit ($TS > 400$). The higher likelihood of the \textit{Fermi} Galactic diffuse model is a result of it being fit in many small energy bins, rather than assuming a single power law over the entire energy range, as we do. A single power law does not capture the curvature in the spectrum. Therefore, we tested a log-parabola and a broken power law spectrum for the \hi\ template and find the fits strongly favored the broken power law spectrum with a $TS$ of at least 200 over the single power law. The \textit{Fermi} Galactic diffuse model still fit better with a $TS \approx 100$. However, for MBM 12, we have enough data to fit several energy bins, more closely replicating the \textit{Fermi} Galactic diffuse model. In every energy bin, our model fits the data as well or better than the standard Galactic diffuse model. Thus, we validate our model based on the \textit{Planck} CO, LAB \hi, and \textit{Planck} dust data.

For the broken power law spectrum, the average energy break occurs at $1.19\pm0.10$ GeV and the power law index changes from $-1.80\pm0.16$ to $-2.82\pm0.06$. The break at around 1 GeV is expected from cosmic ray proton interactions with the cloud and the power law index above the break energy agrees with predictions for gamma-ray emission from a cloud located far from any cosmic ray acceleration site \citep{Aharonian2004}. Changing the spectral shape from a single power law to broken power law results in a 30\% increase in the \hi\ emissivity but does not affect the other gas emissivities. 

We leave the CO and \Avres\ spectra as single power laws for simplicity because \hi\ is the dominant source of gamma-rays across the ROI. Changing the spectrum of the \hi\, CO, and \Avres\ templates to a broken power law resulted in no improvement ($TS < 10$ for six additional free parameters) over the model where only the \hi\ had a broken power law spectrum.

\subsection{MBM 12}

MBM 12 is the best studied high latitude, translucent molecular cloud \citep[e.g.,][]{Pound1990, Ingalls1994, MS1997, Timmermann1998}. It has the highest peak $W_\mathrm{CO}$ of the sample and was tentatively identified as a blazar with a ``confused" designation in the Fermi 2-year catalog (2FGL J0257.9+2025c). It is also coincident with a radio source from the 3$^\mathrm{rd}$ MIT-Green Bank radio survey (MG3 J025805+2029). MBM 12, therefore, merits further analysis. 

\begin{figure}[h]
	\centering
	\includegraphics[width=\linewidth]{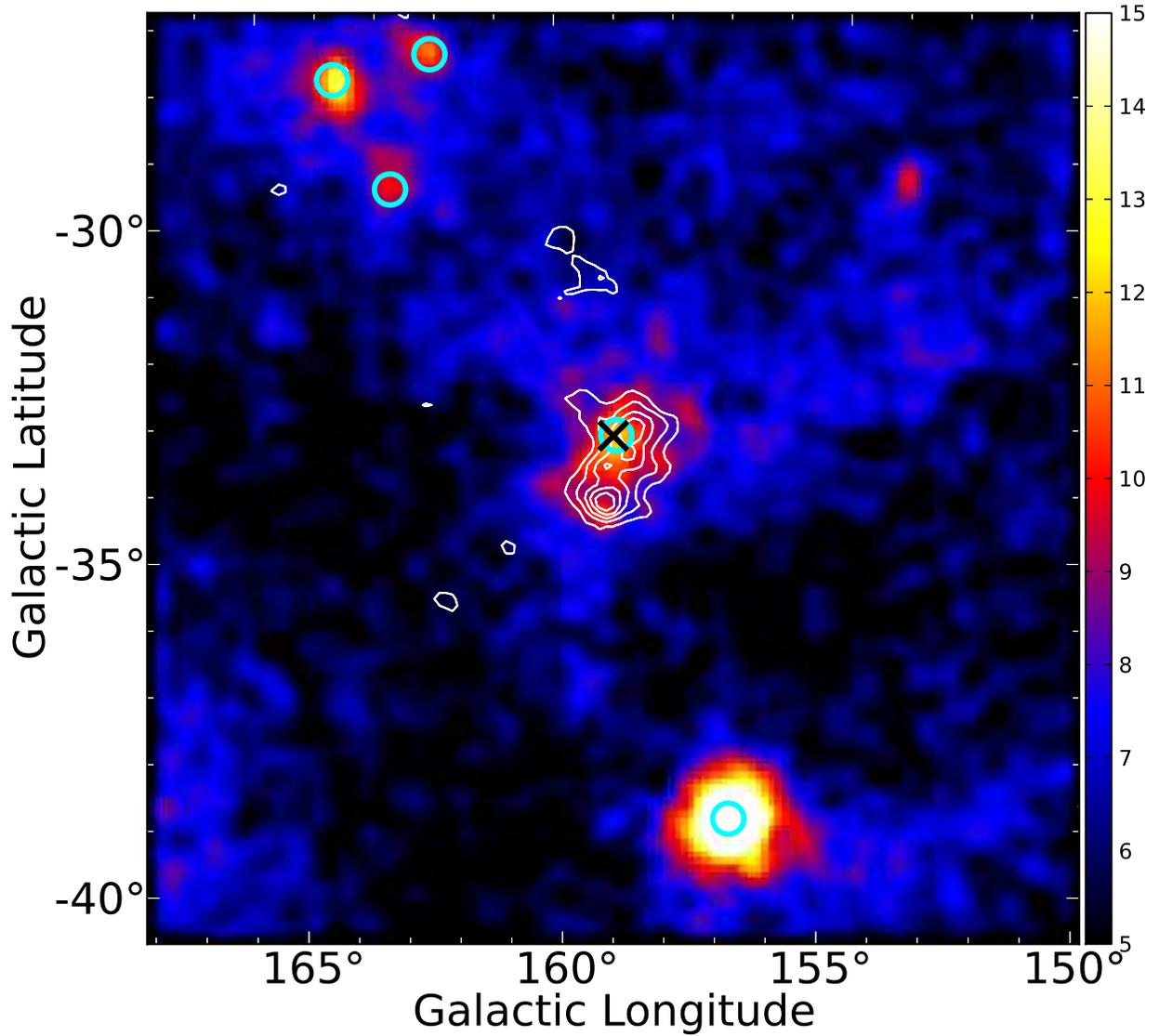}
	\caption{MBM 12 gamma-ray counts map, smoothed with a $\sigma=0.5\degr$ Gaussian, with CO contours at $\WCO = 5, 10, 15, 20,$ and $25$ K \kms. The cyan circles mark the positions of significant 2FGL point sources, and the black $\times$ marks the position of a radio source behind MBM 12. The counts map covers a smaller area than the gas templates in Figure \ref{fig:MBM12temp}.}
	\label{fig:MBM12CMAP}
\end{figure}

Figure \ref{fig:MBM12temp} shows the distribution of the gas in the region around MBM 12. Figure \ref{fig:MBM12CMAP} shows the total number of gamma-rays detected in the region around MBM 12 over the time period analyzed. A 2FGL source, as well as the coincident radio source, lie near the centroid of the molecular cloud.

To establish the molecular cloud origin of the gamma-rays over an AGN source, we perform both a spatial and a variability test. The results of the spatial tests are given as $TS_{\rm AGN}$ and $TS_{\rm ex}$. MBM 12, after removing 2FGL J0257.9+2025c, has $TS_\mathrm{AGN} = 96$ which verifies the presence of the 2FGL point source. All other $TS$ values are calculated with this point source included in the model. A $TS_{\rm ex} = 308$ implies that the gamma-ray emission is extended. The results from these two spatial tests argue in favor of the translucent cloud origin of the gamma-ray emission.

For the variability test, we compare the gamma-ray flux in 15-week time bins to the flux over the entire 58 month period of the survey. As MBM 12 lies far from possible cosmic ray acceleration sites, the cosmic ray source is dominated by the mean steady-state diffusion of cosmic rays through the Galaxy, and should be constant. As a result, significant variability in the gamma-ray flux from MBM 12 would indicate a background AGN, which are the most common gamma-ray sources at high latitude \citep{Fermicat2012} and typically exhibit variability \citep{AckermannAGN}.

\begin{figure}[h]
	\centering
	\includegraphics[width=\linewidth]{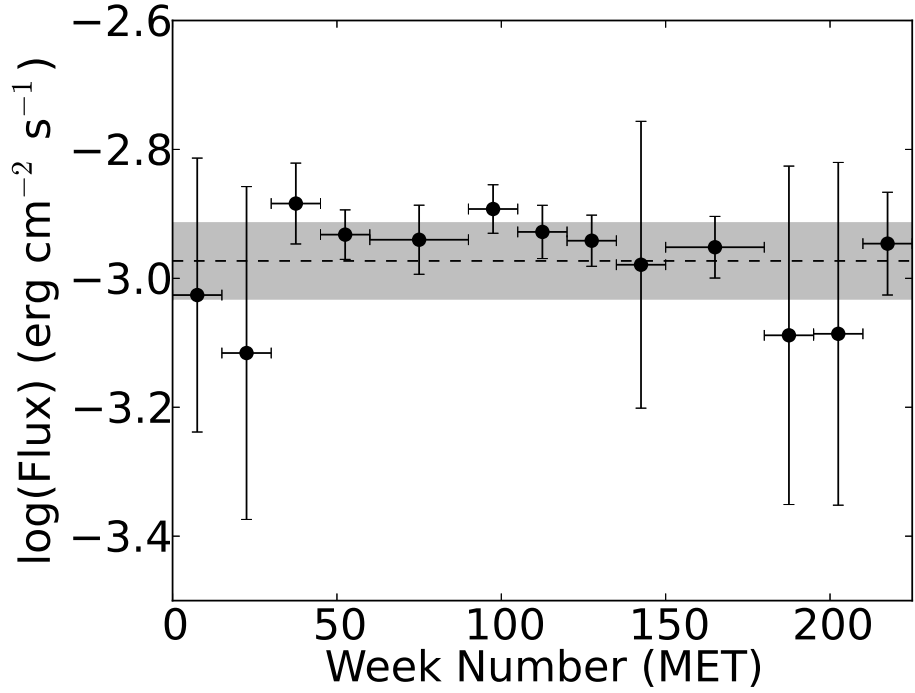}
	\caption{Lightcurve of the total gas flux from the MBM 12 ROI in mission elapsed time (MET). The dashed line is the flux determined by fitting the ROI over the entire time range and the gray shaded region is the statistical uncertainty on this flux.}
	\label{fig:MBM12lc}
\end{figure}

The lightcurve of MBM 12 is given in Figure \ref{fig:MBM12lc} and is the sum of the gamma-ray fluxes from \hi, CO, and dark gas. By eye, they all appear to have constant emission over the observed period. Time bins 4 and 5 as well as 10 and 11 were combined to minimize convergence errors. The dashed line corresponds to the integrated flux found over the entire time period, and the shaded region is the associated uncertainty. With 13 time bins, we find $\chi^2_{13} =1.1$ when comparing the flux to a constant, implying that the gamma-ray flux from MBM 12 is consistent with zero variability.

The tests show that we both detect diffuse gamma-ray emission from MBM 12 and confirm the existence of 2FGL J0257.9+2025c, associated with MG3 J025805+2029. Because we can recover point sources that have been intentionally removed from the model, we gain confidence that we can detect other new sources not included in the 2FGL and disentangle them from diffuse cloud emission. 

As the brightest source in the survey, we also use MBM 12 to test the validity of our chosen energy range. Data analyzed in this region between 10 to 100 GeV shows no significant emission from molecular gas, with a $TS_{\rm \htwo} = 4$. Additionally, the \hi\ emissivity between 250 MeV and 100 GeV differs by only 5\% from the \hi\ emissivity between 250 MeV and 10 GeV. This difference is well within the systematic uncertainty. Extending the analysis to 100 GeV, therefore does not significantly change the emissivity.

\subsection{Uncertainties}\label{sec:uncertain}

We are ultimately trying to probe for small effects, such as possible gradients in the cosmic ray flux in the Solar neighborhood, so it is important to identify and characterize as many sources of uncertainty as possible. Systematic errors introduced by the LAT instrument are estimated to be around 10\% \citep{Fermicat2012}. They arise primarily from uncertainties in the instrument response function, the energy determination of the photons, and the effective area of the LAT. The remaining uncertainties are derived from the results of the likelihood analysis.

Non-local \hi, CO, and dark gas each have two free parameters associated with them and local \hi\ has four free parameters. In addition, the inverse Compton and isotropic emission templates include a normalization term, bringing the total number of template free parameters to 12. Additional, compounded, free parameters are used to create the \Avres\ maps and are discussed below. Point sources add additional free parameters and models with a lot of free parameters tend to converge poorly or force a parameter to one of its limits, which skews the error calculation. Therefore, our analysis procedure removes weak point sources. We test models with different initial values and find that when the models converge, the resulting gamma-ray fluxes and photon indices typically vary by less than 10\%. We take the combined uncertainty due to the LAT systematics and the likelihood variations to be $\sim$15\%.

The uncertainties in the emissivities depend on the detailed model inputs. The LAB survey measured the radiation roughly $-400$ \kms\ to $+400$ \kms\ around the 21 cm line with a sensitivity of 0.09 K \citep{Kaberla2005}. The high velocity gas is all very far away while the low velocity gas is much closer, so we perform a velocity cut of $\pm 20$ \kms on the data to separate nearby gas from far. Most of the \hi\ emission in our clouds lies inside of this cutoff. The velocity cutoff is a fairly small source of error. Generally, the higher velocity gas contributes at most 10\% more column density. High velocity gas, given its much farther distance, is expected to contribute little to the total gamma-ray flux. When the higher velocity \hi\ is significantly detected, the effect on the \hi\ flux of the cloud is $< 10\%$.

In addition, the \hi\ template suffers from uncertainty due to the assumed spin temperature. Spin temperatures likely change across an ROI and even across an individual cloud \citep{Fukui2014MBM}. We tested four spin temperatures: $T_S = 80$ K, 125 K, 400 K, and $T_S\to \infty$. The gamma-ray flux from \hi\ changed less than 7\% while the emissivity decreased by 15\% with increasing $T_S$ over the entire range of $T_S$ values.

Surrounding the molecular cloud may be a shell of optically thick \hi\ with a very low spin temperature, that also contributes to the dark gas phenomenon \citep{Fukui2014DG,Stanimirovic2014}. Any unaccounted for \hi\ due to a lower spin temperature than the 125 K used should be captured in the \Avres\ map. This should not affect the \hi\ emissivity or gamma-ray flux by more than 5\%, as the \hi\ in the cloud accounts for less than 5\% of the total \hi\ emission across the ROI.

The main uncertainties from the CO template comes from the 2$\sigma$ cutoff used to remove noise, and the $^{13}$CO contamination of the \textit{Planck} \WCO data \citep{Ade2013CO}. We tested a 1$\sigma$ cutoff as well and saw emissivities and fluxes systematically lower by around 5\%. This is expected since the CO covers a larger area in the 1$\sigma$ template versus the 2$\sigma$ template. Yet the distribution of CO does not change much; most of the difference between the 1$\sigma$ and 2$\sigma$ templates is noise and distributed roughly uniformly around the map. Nearly the same flux is being emitted from a larger amount of CO, therefore the emissivity decreases accordingly.

The \Avres\ map is a linear combination of the LAB HI map, the \textit{Planck} CO map, and the \citet{Abergel2014} color excess map, where the uncertainties for the color excess are generally less than 8\%, with an average across each ROI of less than 4\%. The uncertainty in the color excess dominates the \Avres errors over those of the \hi\ and CO maps. The fluxes and emissivities for CO and \Avres\ are insensitive to the change in spin temperature. Even between the two extremes, $T_S = 80$ K and $T_S \to \infty$, the \Avres\ template changes by less than 2\% on average, which is much smaller than the uncertainties of the color excess map. 

Additional uncertainties in the \Avres\ map arise in $R_V = \Av/\EBV$. While overall variations in $R_V$ will not affect the \hi\ emissivity, CO and dark gas emissivities may be affected if $R_V$ varies across the ROI. $R_V=3.1$ in diffuse regions \citep{Cardelli1989} but increases to $R_V \ge 4$ in some molecular clouds \citep{Vrba1993,Kandori2003}. A constant $R_V$ may underpredict the extinction values in molecular clouds by as much as 30\%. The uncertaintiy in the emissivity is not clear due to the relationship between \Avres\ and \WCO.

To study this effect, other dust templates may be more approprate. For example, a properly scaled $\tau_{353}$ map is recommended for regions of higher \Av\ \citep{Abergel2014} instead of the \EBV map. This test was performed in the Chamaeleon region in \citet{Ade2014}. Changing the dust map necessarily affects the \Avres\ map, and therefore also the fitted and subsequently calculated quantities. We test both dust maps in MBM 12. Comparing both models yields a $TS = 2$. Therefore neither \EBV\ nor $\tau_{353}$ are preferred over the other. The \Avres\ emissivity changed by up to 30\% while the \hi\ and CO emissivities changed by less than 5\%. We quantify the effect of the dust template more thoroughly in the full survey and defer that discussion to future work. We argue that with only 9 clouds in our current sample, a 30\% change in dark gas emissivity does not significantly affect the results discussed in Section \ref{sec:disc}.

Finally, we also tested a model using the dust map as the sole gas tracer. In principle, the extinction should trace all gas species and thus should provide a model with significantly fewer free parameters. The dust model, however, does not reproduce the gamma-ray data. The $TS$ between model \textit{CODG} and the model with dust alone is 175; the dust model fits significantly worse than our baseline model, and exhibits large, structured residuals in regions with CO and dark gas. This may be due to varying dust properties \citep{Ade2014}, a lack of sensitivity in dense regions, a limited range of applicablility, or a nonlinear response. This validates our combination of gas templates. 

In sum, there a is 15\% systematic uncertainty in the gamma-ray flux from both the LAT instrument and the likelihood analysis along with an additional uncertainties in the emissivities due to the choice of model of $q_\mathrm{\hi}{}^{+17\%}_{-8\%}$, $q_\mathrm{CO} \pm 5\%$. For this study, we adopt $q_\mathrm{\Avres}\pm 8\%$.

\section{Discussion}\label{sec:disc}

From these observations, we may derive the gamma-ray emissivity, the gamma-ray photon flux per H-atom, for the brightest clouds within 270 pc of the Sun between 250 MeV and 10 GeV. The fit treats the energy range as a single bin. Table \ref{tab:fit} lists the emissivities for \hi, CO, and the dark gas, while Table \ref{tab:fit_ICiso} lists the remaining model parameters excluding point source normalizations. The \hi\ emissivities are all generally consistent with each other, and close to the average emissivity of $q_{\rm \hi}(250\mathrm{\;MeV} - 10\mathrm{\;GeV}) = (8.1\pm1.4)$\ee{-27} photons \cmtwo\ \pers\ sr$^{-1}$ H-atom$^{-1}$. This result is consistent with the emissivities, all between 250 MeV and 10 GeV, found in the Cygnus region, $(8.76\pm0.33)\times 10^{-27}$ \citep{Ackermann2012Cyg}, and that found in the region around the  Cepheus/Polaris Flare, $(9.2\pm0.3)\times 10^{-27}$ \cite{Ackermann2012c}. The Chamaeleon region has been analyzed twice and the \hi\ emissivity above 250 MeV was found to be $(7.2\pm0.1)\times 10^{-27}$\cite{Ackermann2012c} and $(10.8\pm0.4)\times 10^{-27}$\citep{Ade2014} and the \hi\ emissivity around R Coronae Australis was found to be $(10.2\pm0.4)\times 10^{-27}$ \citep{Ackermann2012c}. All values are less than 2$\sigma$ from our results.

The emissivities for the molecular gas tracers, the CO and dark gas, vary more significantly. The CO component has an average emissivity of $q_{\rm CO}(250$ MeV -- 10 GeV$) = (1.6\pm0.6)$\ee{-6} photons \cmtwo\ s$^{-1}$ sr$^{-1}$ (K \kms)$^{-1}$ and the dark gas component has an average emissivity of $q_{\rm \Avres}(250$ MeV -- 10 GeV$)=(1.5\pm0.7)\times10^{-5}$ photons \cmtwo\ \pers\ sr$^{-1}$ mag$^{-1}$, taking into account statistical errors. The CO emissivity was found to be $(3.01\pm0.16)\times 10^{-6}$ in the Cygnus region while the clouds Chamaeleon, R Coronae Australis, and Cepheus/Polaris flare have $(1.04\pm0.08)\times 10^{-6}$, $(1.9\pm0.2)\times 10^{-6}$, and $(1.23\pm0.05)\times 10^{-6}$ photons \cmtwo\ \pers\ sr$^{-1}$ (K \kms)$^{-1}$, respectively \citep{Ackermann2012c}. The dark gas emissivity for all four regions was $(2.75\pm0.26)\times 10^{-5}$ \citep{Ackermann2012Cyg} and $(1.36\pm0.04)\times 10^{-5}$, $(2.3\pm0.2)\times 10^{-5}$, and $(1.38\pm0.08)\times 10^{-5}$ photons \cmtwo\ \pers\ sr$^{-1}$ mag$^{-1}$ \citep{Ackermann2012c}, respectively. Our results are entirely consistent with the previously found values.

\subsection{Cosmic Rays in the Solar Neighborhood} \label{sec:CR}

\begin{figure*}[h]
\centering
\begin{minipage}[t]{0.45\textwidth}
	\includegraphics[width=\linewidth]{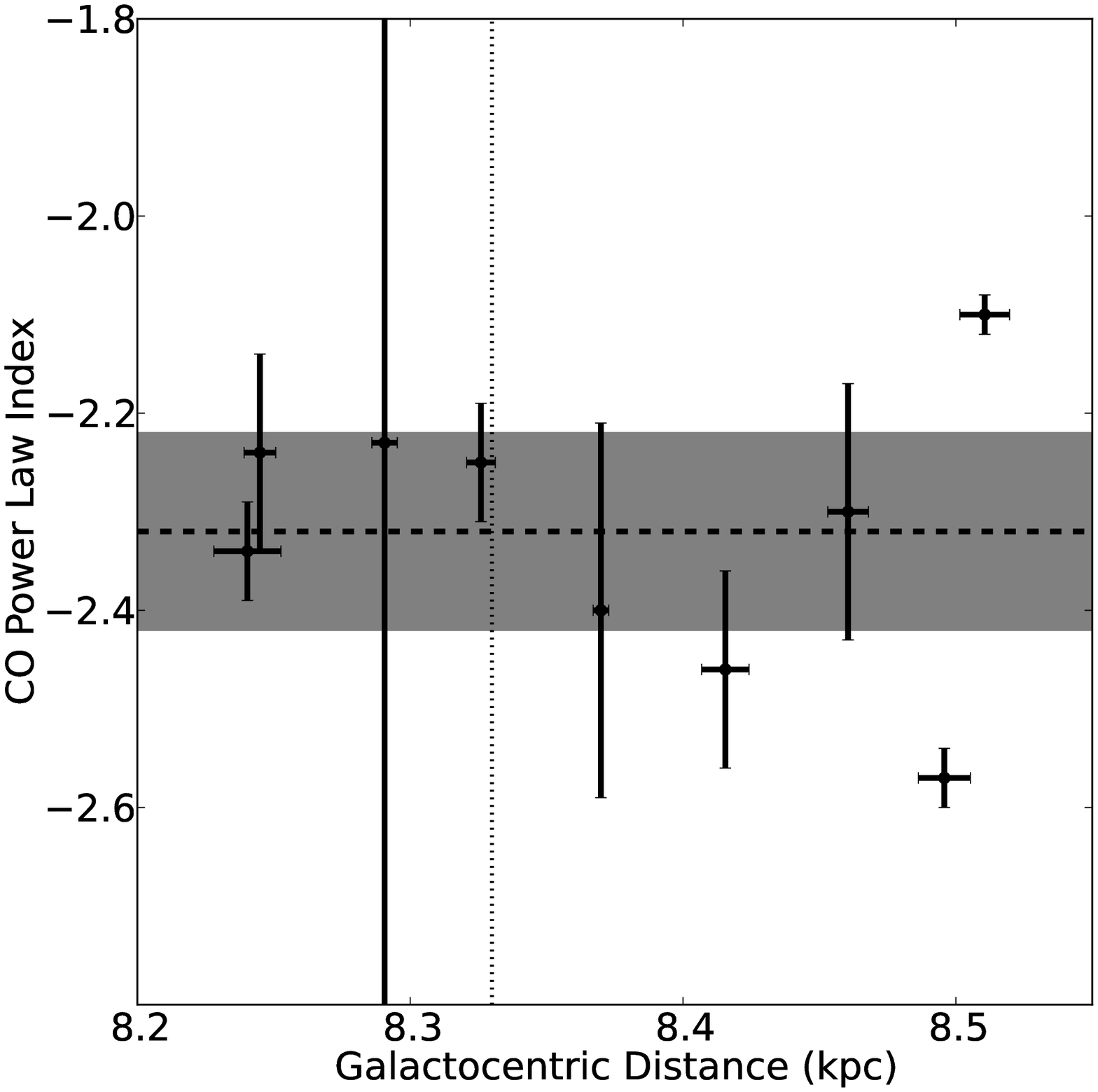}
\end{minipage}
\hspace{\fill}
\begin{minipage}[t]{0.45\textwidth}
	\includegraphics[width=\linewidth]{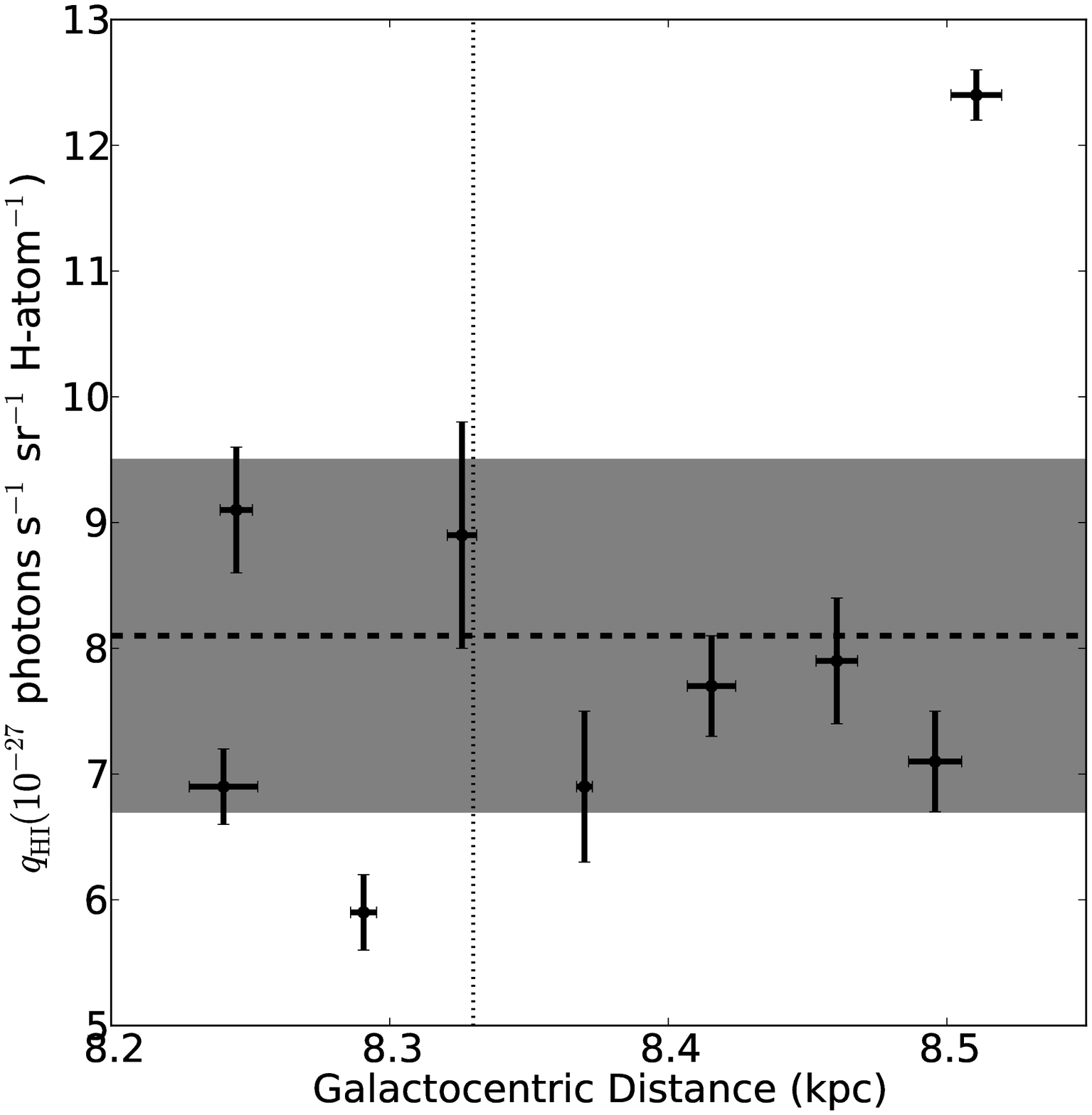}
\end{minipage}
	\caption{Power law index for gamma-ray emission from CO (left) and \hi\ emissivity (right) as a function of Galactocentric distance, with the distance to the Galactic Center $R_0 = 8.33 \pm 0.35$ \citep{Gillessen2009}, marked with a vertical dotted line. The horizontal dashed lines are the best fit constant power law index and \hi\ emissivity, respectively, and their uncertainty is the shaded region.}
	\label{fig:Rvqhi}
\end{figure*}

A change in the gamma-ray emissivity indicates a change in the incident cosmic ray flux. On the right of Figure \ref{fig:Rvqhi}, we plot the \hi\ emissivity and see no variation. MBM 12 has a high value, though systematic uncertainties place it within the 2$\sigma$ range of the average \hi\ emissivity. A constant \hi\ emissivity is in agreement with the conclusions of \citet{Ackermann2012c} and the predictions of GALPROP models \citep{Strong2004}. \citet{Abdo2010} looked for gradients in $q_\mathrm{\hi}$ and compared it to predictions from GALPROP. They measure a 10\% decrease in $q_\mathrm{\hi}$ from the Gould Belt to the local arm, a distance of about 1 kpc. The maximum Galactocentric distance between any of the clouds in this survey covers about 0.25 kpc, so we might expect to see a 2.5\% change in $q_\mathrm{\hi}$ across our sample, which is well within our quoted uncertainties. 

Additionally, we report the gamma-ray photon index from the CO to isolate the molecular cloud. The indices from CO and dark gas components are identical within the uncertainties. This index is similar to the average of the two power law indices in the broken power law spectrum of \hi. The indices of the clouds are listed in table \ref{tab:fit} and are taken from the CO fit. Where CO is not detected, the index is taken from the dark gas fit. The left side of Figure \ref{fig:Rvqhi} shows the power law index for every cloud. They all lie close to the best fit value $-2.25 \pm 0.10$ with no evidence of a variation. The lack of detected variations implies a uniform cosmic ray flux incident on each cloud verifying that there are no cosmic ray sources near the clouds studied. The farthest a cloud can be from a supernova remnant and still receive a cosmic ray excess is 100-200 pc for a $10^4$ year old supernova remnant \citep{Gabici2011}.

\subsection{X-factors}\label{sec:Xfac}

We trace molecular gas with two components, \WCO\ and \Avres, and so we require two conversion factors to estimate the column density of \htwo. \Avres\ is not entirely molecular gas, however; some fraction is atomic hydrogen \citep{Fukui2014MBM, Stanimirovic2014, Fukui2014DG}. The molecular gas column density can therefore be written as:

\begin{equation}\label{eq:x-def}
	N(\htwo) = \xco\WCO + f\xav\Avres,
\end{equation}

\noindent where $f$ is the molecular fraction of the dark gas. This expression is proportional to the traditional X-factor, $X_\mathrm{CO}$, where CO is assumed to trace all the \htwo.

Analysis of gamma-rays alone does not determine what fraction of dark gas is \htwo, but we can put upper and lower limits on $X_{\rm CO}$:

\begin{equation}\label{eq:xco}
	X_{\rm CO} = \xco + \xav \left(\frac{f\Avres}{\WCO}\right).
\end{equation}

Assuming a constant cosmic ray flux and that cosmic rays penetrate the entire cloud, every proton should be subject to the same number of cosmic ray interactions. This assumption was verified in the Chamaeleon clouds in \citet{Ade2014}, and implies $q_\mathrm{\mathrm{H}_2} = 2 q_\mathrm{\hi}$. The gamma-ray emission is proportional to the number of molecules, but we use \WCO\ in our analysis: $q_\mathrm{\mathrm{H}_2}N(\htwo) = q_\mathrm{CO} \WCO$. This leads to the relationship $\xco =q_\mathrm{CO}/2 q_\mathrm{\hi}$. A similar argument leads to the relationship $\xav = q_\mathrm{\Avres}/([1+f]q_\mathrm{\hi})$, where the factor $[1+f]$ arises because a fraction of the gas represented by molecular hydrogen which has two protons. 

\begin{figure}[h]
	\centering
	\includegraphics[width=\linewidth]{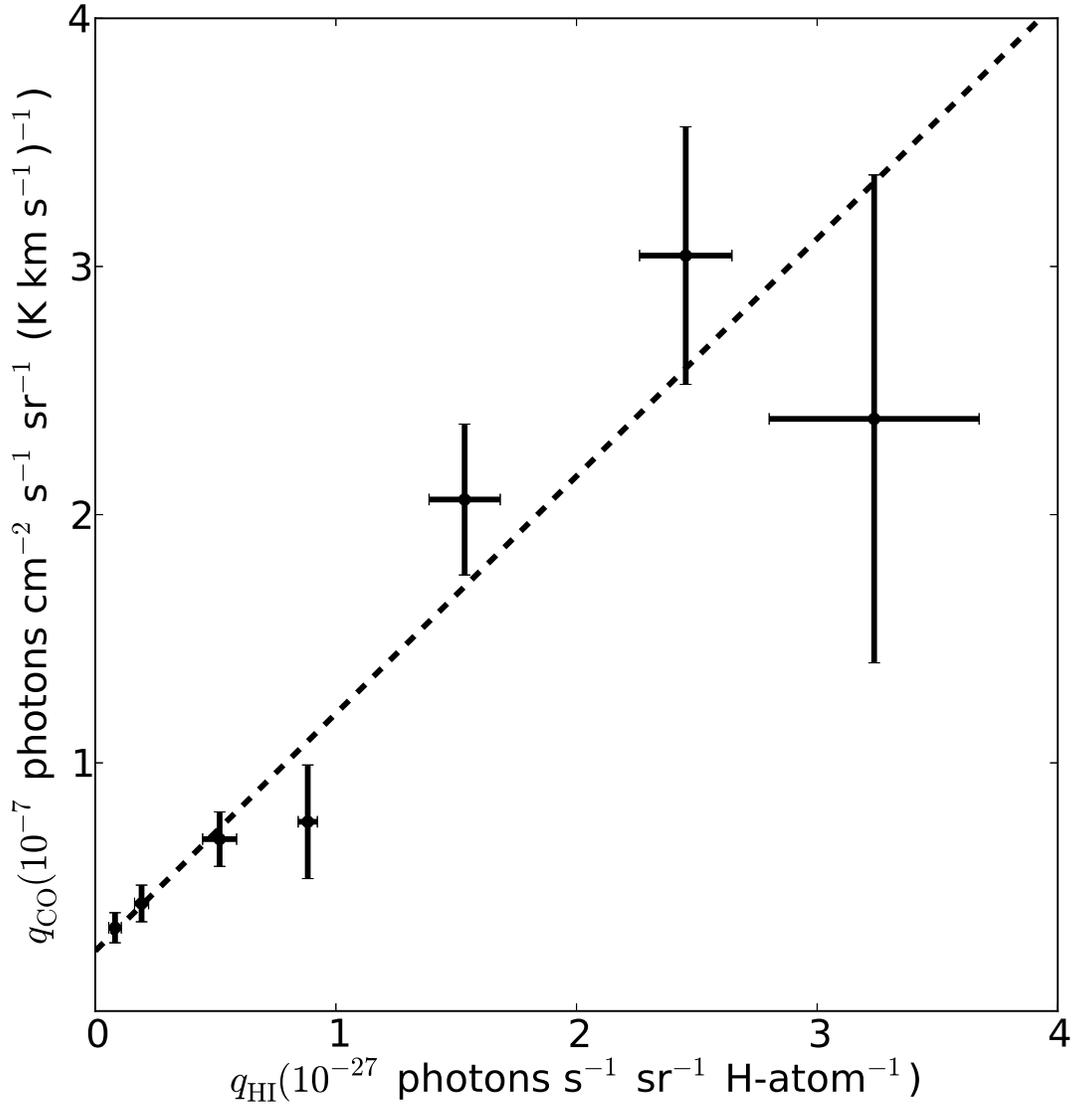}
	\caption{Gamma-ray emissivity for CO versus emissivity for \hi\ in MBM 12 shows a linear relationship.}
	\label{fig:MBM12covhi}
\end{figure}

\begin{figure}[h]
	\centering
	\includegraphics[width=\linewidth]{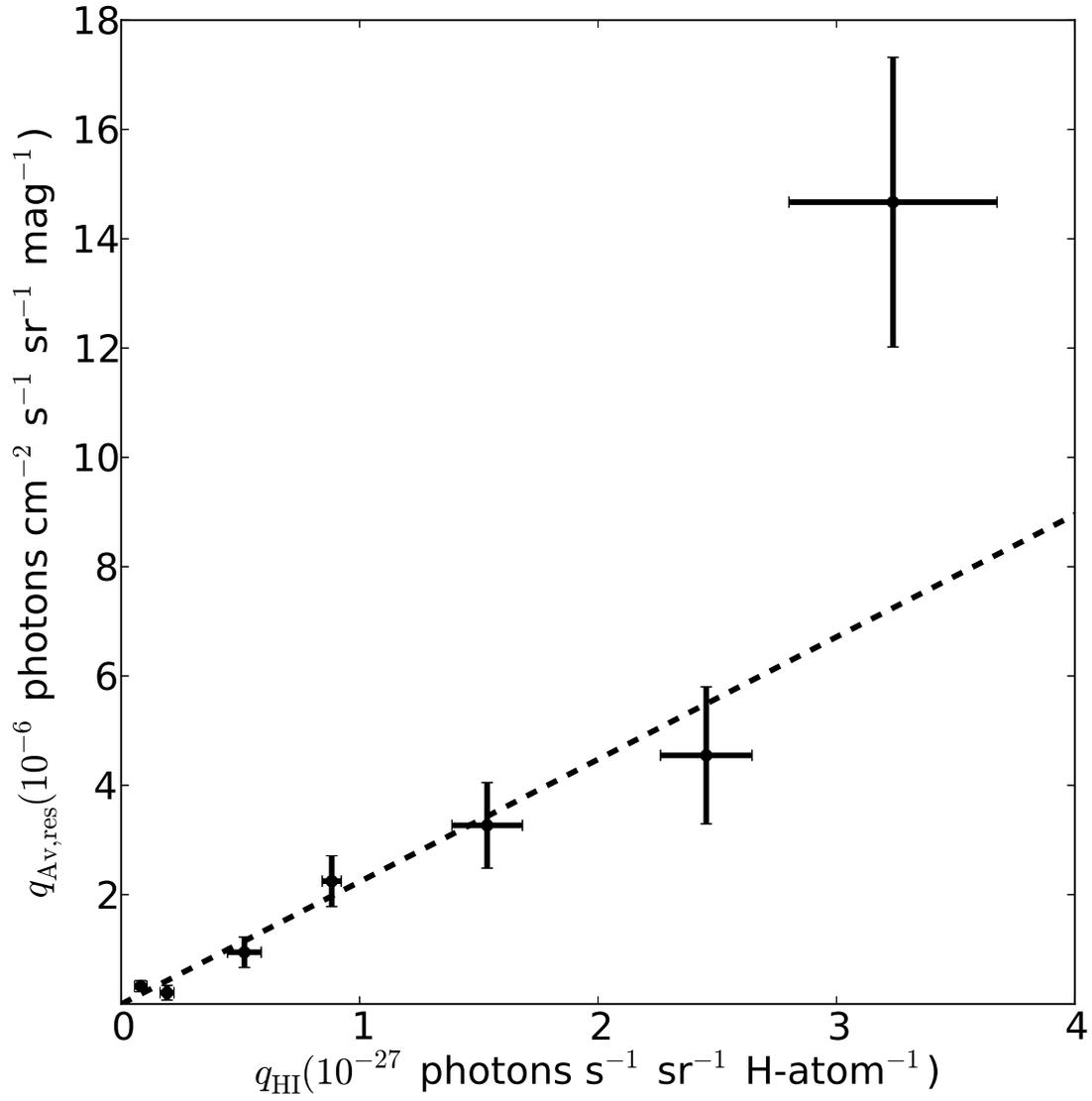}
	\caption{Gamma-ray emissivity for dark gas versus emissivity for \hi\ in MBM 12 shows a linear relationship.}
	\label{fig:MBM12dgvhi}
\end{figure}

From the emissivities of MBM 12 in different energy bins given in Table \ref{tab:fit_energy}, we plot $q_{\rm CO}$ versus $q_{\rm \hi}$ in each energy bin in Figure \ref{fig:MBM12covhi}. We find a linear relationship between the two emissivities and plot the best fit line. Similarly, we plot $q_{\Avres}$ versus $q_{\rm\hi}$ in Figure \ref{fig:MBM12dgvhi}. Again, we see a linear relationship, which gives us confidence that the cosmic ray flux at MBM 12 is constant and penetrates through the entire cloud. For MBM 12, the slope of the best fit line in Figure \ref{fig:MBM12covhi} is $\xco = (4.8\pm 1.2)\ee{19}$ \cmtwo\ (K \kms)$^{-1}$, which agrees with the value in Table \ref{tab:calc} obtained as a ratio of the emissivities from Table \ref{tab:fit}. The slope of the best fit line in Figure \ref{fig:MBM12dgvhi} is $\xav = (22.4\pm3.5)\ee{20}$ \cmtwo\ mag$^{-1}$, roughly $2.5\sigma$ from the value in Table \ref{tab:calc}.

Most clouds are not detected significantly enough to make a spectrum, so we cannot verify the linear relationship in each case. MBM 12 has the highest \Av\ of any of the clouds in this study, so we conclude cosmic rays penetrate through every cloud in this study. That our clouds are far from cosmic ray acceleration sources and the \hi\ emissivity is constant across them gives us some confidence that the cosmic ray flux is constant across an entire molecular cloud. We therefore assume $\xco =q_\mathrm{CO}/2 q_\mathrm{\hi}$ and $\xav = q_\mathrm{\Avres}/([1+f]q_\mathrm{\hi})$ for every cloud in this study.

The average value for \xco\ among the clouds is ($1.1\pm0.4$)\ee{20} \cmtwo\ (K \kms)$^{-1}$. This is consistent with previous gamma-ray studies of nearby molecular clouds \citep{Ackermann2012c}, where $X_{\rm CO} \la 1\ee{20}$ \cmtwo\ (K \kms)$^{-1}$. This value is lower than that found for high latitude clouds of $1.67\ee{20}$ \cmtwo\ (K \kms)$^{-1}$ \citep{Paradis2012}. Our \xco\ for Cham-East {\small II} is consistent with a recent analysis of the Chamaeleon cloud complex \citep{Ade2014} which finds $X_\mathrm{CO} \sim 7\ee{19}$ K \kms. However, \xav\ is always higher than \xco\ as seen in Table \ref{tab:calc}, with an average of ($19.0\pm8.8$)\ee{20}. As long as any fraction of \Avres\ represents \htwo, the combination of the two will increase $X_\mathrm{CO}$. 

Assuming $f=1$, we report the average $X_\mathrm{CO}$ of each cloud in Table \ref{tab:calc}. The average among all the clouds is $X_\mathrm{CO} = (1.6 \pm 0.5)$\ee{20} \cmtwo\ (K \kms)$^{-1}$ with large cloud-to-cloud variations. This result agrees with previous studies of $X_\mathrm{CO}$ at high latitudes \citep{Magnani1995,Paradis2012}, suggesting the dark gas in these previous studies is primarily molecular. Our average $X_\mathrm{CO}$ is higher than the average value found in Perseus \citep{Lee2014}, which may suggest that the molecular fraction of the dark gas in Perseus is significantly lower than 1 or cosmic rays do not penetrate deeply into the CO-bright regions of the cloud. Our $X_\mathrm{CO}$ for every cloud is consistent with that found in the high latitude cloud MBM 40 which found an average $X_{\rm CO} = 1.3\ee{20}$ \cmtwo\ (K \kms)$^{-1}$ \citep{Cotten2013}. 

\begin{figure}[h]
	\centering
	\includegraphics[width=\linewidth]{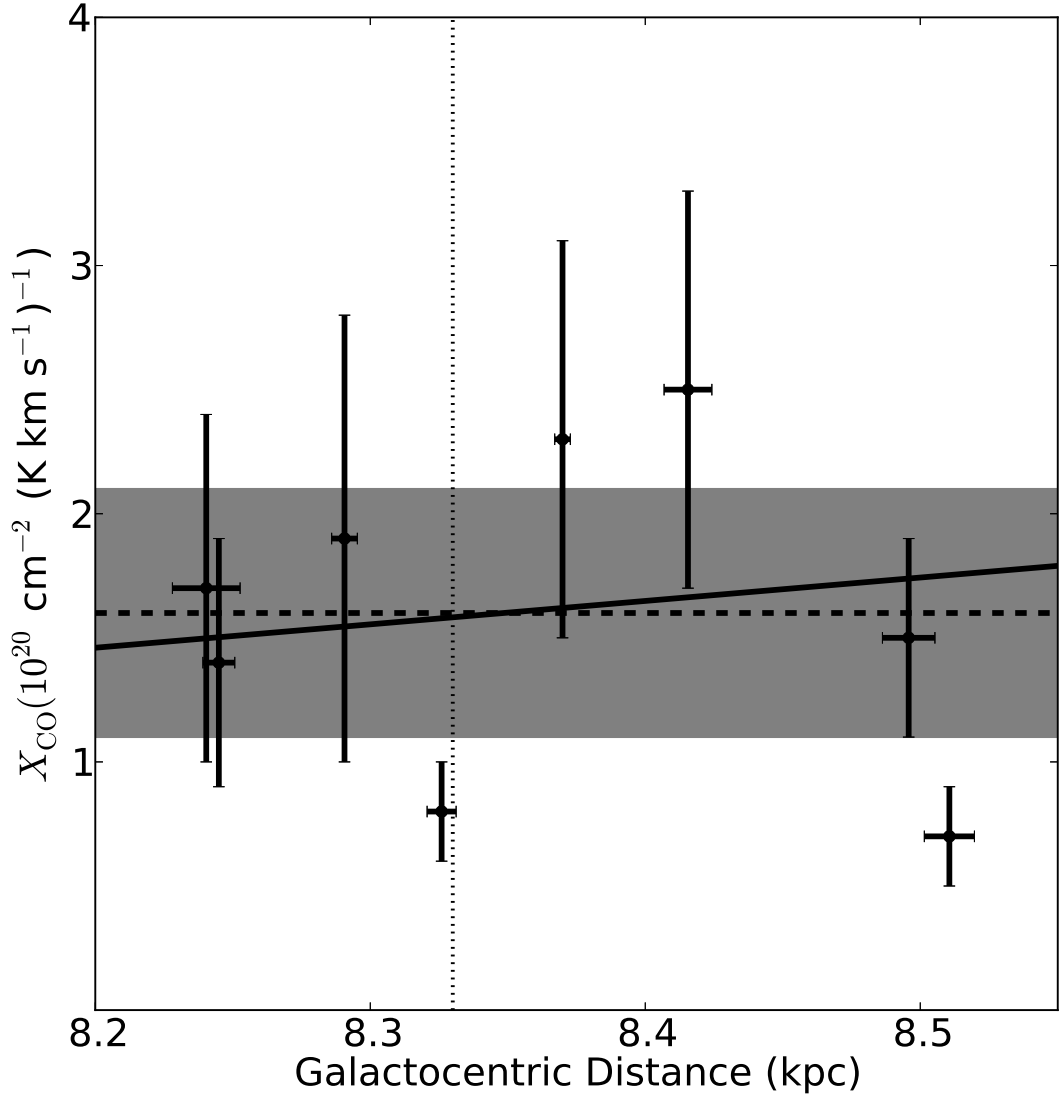}
	\caption{$X_\mathrm{CO}$ as a function of the distance from the Galactic Center, as in Figure \ref{fig:Rvqhi}. The solid line represents a variation in $X_\mathrm{CO}$ due to metallicity from \citet{Israel2000}, rescaled according to \citet{Pineda2013}.}
	\label{fig:Xco_dist}
\end{figure}

Figure \ref{fig:Xco_dist} shows $X_\mathrm{CO}$ as a function of Galactocentric distance. The solid line in the figure shows one of the more extreme variations of $X_\mathrm{CO}$ considered near the Solar neighborhood \citep{Israel2000}. As with the gamma-ray spectrum, we find no evidence for overall variation in $X_\mathrm{CO}$ over this small extent of 270 pc.
% Perseus: also the usual cloud-to-cloud differences are a factor of a few, so...

\begin{figure}[h]
	\centering
	\includegraphics[width=\linewidth]{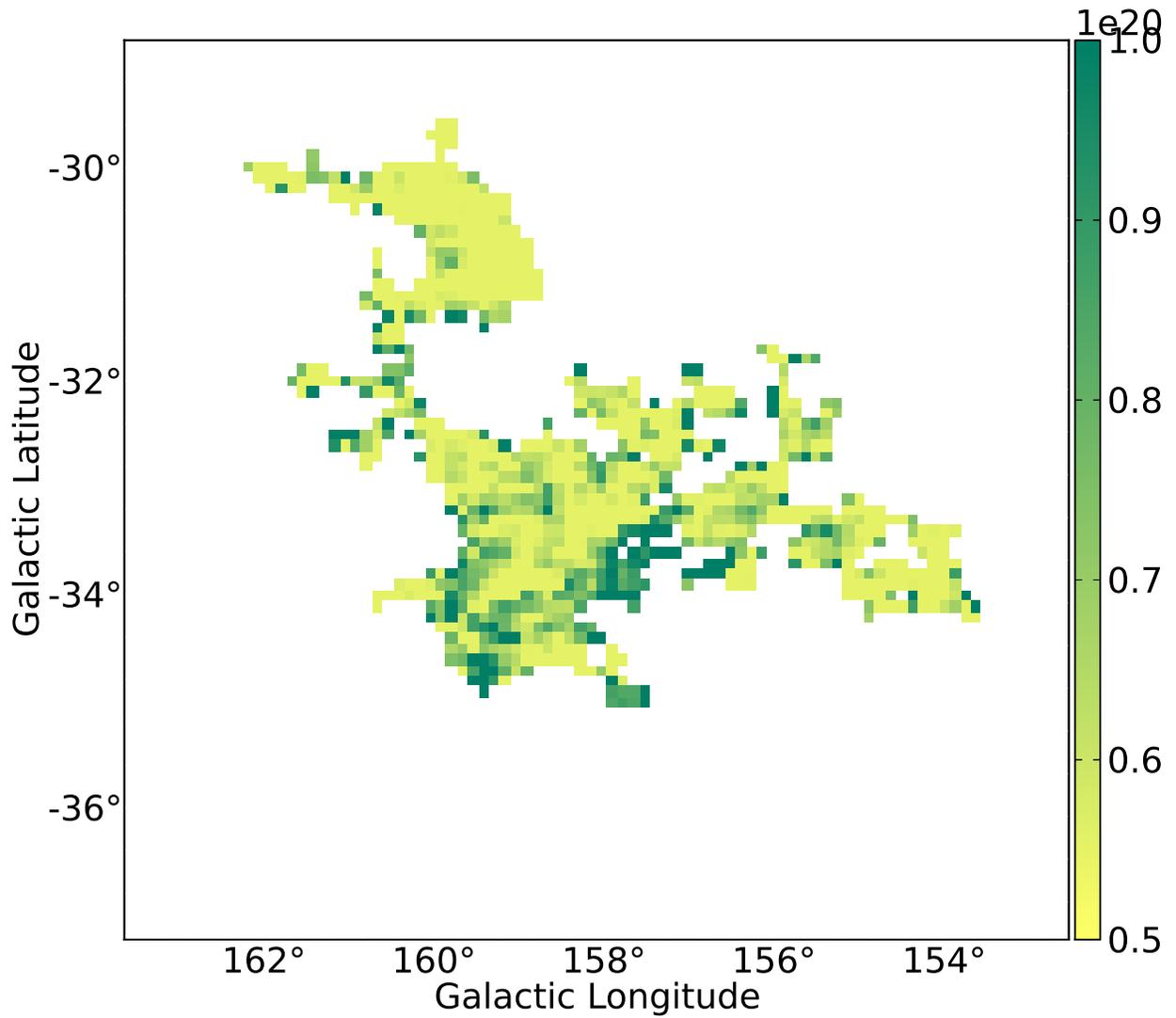}
	\caption{A map of $X_\mathrm{CO}$ from Equation \ref{eq:xco} assuming $f=1$, across the CO-emitting portion of MBM 12  in units of $10^{20}$ \cmtwo\ (K \kms)$^{-1}$.}
	\label{fig:Xco_comp}
\end{figure}

Figure \ref{fig:Xco_comp} shows $X_\mathrm{CO}$ across the CO-bright part of MBM 12, assuming $f=1$ uniformly across the entire cloud. This places an upper limit on the value of $X_\mathrm{CO}$. $X_{\rm CO}$ appears to increase towards the edge of the CO-emitting region where the total extinction drops by as much as an order of magnitude. The middle of the cloud shows a low $X_{\rm CO}$, where the total extinction increases due to increased gas density. A higher density supports the transition to fully molecular gas. This transition reduces the fraction of dark gas in the cloud. Thus, the departure of $X_\mathrm{CO}$ from \xco\ in the CO-bright part of the cloud may reflect the transition of the cloud from atomic to molecular gas or it may reflect a level of small-scale clumpiness in the cloud, supporting some interior dissociation along the line of sight. If $f$ decreases to zero toward the edge of the CO-bright part of the cloud, we can potentially recover a constant $X_\mathrm{CO}$ across the cloud, though not expected to occur \citep{Wolfire2010}.

As can be seen by visual inspection of the CO and dark gas maps in Figure \ref{fig:MBM12temp}, dark gas extends beyond the CO boundaries as expected from numerical studies \citep{Wolfire2010,Velusamy2010}. In these regions, we can only place limits on $X_\mathrm{CO}$. Therefore, we do not attempt to estimate $X_\mathrm{CO}$ beyond the CO-bright boundary of the molecular cloud.

Finally, as an initial evaluation of the dust model, we replace the \textit{Planck} \EBV\ map in MBM 12 with the \textit{Planck} $\tau_{353}$ dust opacity map, scaled by the given ratio: $\EBV/\tau_{353} = 1.49\times 10^{4}$ \citep{Abergel2014}. In MBM 12, changing the dust model greatly affects \xav. However $X_{\rm CO}$ changes by $\pm 0.5$\ee{20} \cmtwo\ (K \kms)$^{-1}$, which is smaller than the variations of $X_{\rm CO}$. $X_{\rm CO}$ is thus moderately insensitive to large variations in \xav\ due to suppression by the ratio $\Avres/\WCO$, which averages between 0.01 -- 0.10 mag (K \kms)$^{-1}$. It is worth noting that, while the distribution of $X_\mathrm{CO}$ seen in Figure \ref{fig:Xco_comp} depends on the dust tracer used, the magnitude of the variation of $X_{\rm CO}$ is larger than any uncertainty of $X_{\rm CO}$. We will quantify the uncertainty due to changing the dust tracer in future work. 

\section{Conclusions}

We study the gamma-ray emission from nine high latitude, translucent molecular clouds and find the gamma-ray spectrum and emissivity from the gas in the clouds. All nine \textit{Planck}-selected CO clouds were significantly detected, showing extended emission associated with molecular gas. We estimate the systematic uncertainties associated with modeling the gamma-ray emission with \hi, CO, and dark gas templates. The \hi\ emissivity does not vary across the regions in the sample. For some clouds, the gamma-ray emission is dominated by the CO-emitting gas, while for other clouds the CO-dark gas dominates the gamma-ray emission. % Where do we say thing????

\xco\ ranges from 0.2\ee{20} \cmtwo\ (K \kms)$^{-1}$ to 1.6\ee{20} \cmtwo\ (K \kms)$^{-1}$ with large uncertainties, and \xav\ spans a similarly large range around $10^{21}$ \cmtwo\ mag$^{-1}$. In order to compare these X-factors to the traditional conversion factor between \WCO\ and \htwo\, \xco\ and \xav\ are added together assuming the molecular fraction of the dark gas is 1. The results are consistent with previous studies and suggest no change in cosmic ray flux across $\sim 300$ pc region around the Solar System. The combination of \xco\ and \xav\ may also explain the low values of $X_\mathrm{CO}$ found in \citet{Ackermann2012c} compared with estimates using other methods \citep{Bolatto2013}. However, the choice of dark gas tracer should be made carefully.

\acknowledgments

The authors thank the \textit{Fermi} LAT team for their support, particularly at the 2013 \textit{Fermi} Summer School and through the \textit{Fermi} Science Support Center website, as well as the anonymous referee for insightful comments. This work was supported in part by the NASA New York Space Grant Consortium based at Cornell University (\# NNX10AI94H) and by grant \# 63388-00 41 from the Professional Staff Congress of the City University of New York. TADP acknowledges support from NSF grant AST-1153335.

{\it Facilities:} \facility{\textit{Fermi} (LAT)}

%
%
% Now the figures
%
%

\clearpage

\clearpage

\clearpage

\clearpage

\clearpage

\clearpage

\clearpage

\clearpage

\clearpage

\clearpage

\clearpage

%
%
% Here are the tables
%
%

\clearpage

\begin{deluxetable}{lcccccc}
\tablewidth{0pt}
\tablecaption{Cloud Properties}
\tablehead{
Name		&
$l$		 	& 
$b$		 	&
Peak \WCO 	&
Distance		&
Area \cr
			&
(deg) 		&
(deg) 		&
(K \kms) 	&
(pc) 		&
(deg$^2$) }
\startdata
MBM 36\tablenotemark{a}	 	& 3.9 	& 35.7 	& $27.85\pm 0.74$  & $105\pm 7$ & $4.04^{+1.46}_{-0.90}$\cr 
DIR 071-43 				 	& 71.6	& -42.8 	& $3.69\pm 0.70$ & $170 \pm 20$\tablenotemark{b} & 1.35$^{+1.6}_{-0.85}$\cr 
MBM 55\tablenotemark{c}	 	& 88.5 	& -41.2 	& $15.75\pm 0.70$ & $206^{+8}_{-6}$ & 21.23$^{+18.82}_{-1.75}$ \cr 
MBM 02 					 	& 108.3 	& -51.9 	& $7.45\pm 0.73$ & $206^{+14}_{-12}$& 1.77$^{+0.78}_{-1.64}$\cr 
MBM 04\tablenotemark{d}	 	& 132.9 	& -45.6 	& $15.31\pm 0.78$ & $269^{+16}_{-14}$ & 1.16$^{+9.6}_{-0.39}$\cr 
MBM 32\tablenotemark{e}	 	& 147.6 	& 40.7 	& $8.06\pm 0.62$ & $259^{+14}_{-15}$ & 2.36$^{+4.83}_{-0.79}$ \cr 
MBM 12\tablenotemark{f} 	 	& 159.1 	& -34.3 	& $29.78\pm 0.73$ & $234^{+11}_{-10}$ & 10.24$^{+6.61}_{-3.15}$\cr 
MBM 20\tablenotemark{g} 	 	& 210.9 	& -36.5 	& $12.39\pm 0.72$ & $124^{+11}_{-14}$ & 2.48$^{+1.84}_{-0.71}$\cr
Cham-East {\small II}\tablenotemark{h} & 312.9 	& -28.6 	& $20.25\pm 0.53$ & $150 \pm 20$\tablenotemark{b} & 12.45$^{+4.09}_{-1.85}$\cr
Control						& 250.0	& 30.0	& $< 2.46\pm0.7$&	---			& $<0.06$ \cr
\enddata
\tablecomments{Regions around the \WCO peak included in the gamma-ray analysis include additional molecular clouds. Distances derived from \citet{Schlafly2014} unless otherwise noted.}
\tablenotetext{a}{ROI also includes: MBM 33 through MBM 35, MBM 37, and MBM 38}
\tablenotetext{b}{Distances derived from \citet{Lallement2013}}
\tablenotetext{c}{ROI also includes: MBM 53 and MBM 54}
\tablenotetext{d}{ROI also includes: MBM 03, DIR 121-45}
\tablenotetext{e}{ROI also includes: MBM 27 through MBM 31, HSVMT 24, HSVMT 27, and HSVMT 28}
\tablenotetext{f}{ROI also includes: MBM 07 through MBM 09, and MBM 11}
\tablenotetext{g}{ROI also includes: DIR 203-32}
\tablenotetext{h}{ROI also includes: Cham-East {\small I}}
\label{tab:prop}
\end{deluxetable}

\clearpage

\begin{deluxetable}{lcccccc}
\tablewidth{0pt}
\tablecaption{Coincident Point Sources}
\tablehead{
Cloud Name	&
Fermi Catalog	& 
Power Law	&
Variability		&
Classification 	&
Flags   \cr
			&
Name 		&
Index 		&
Index\tablenotemark{a}		&
		 	&
$(N)$	 }
\startdata
MBM 12 		 & 2FGL J0257.9+2025c & -2.19 & 25.55  &blazar\tablenotemark{b} 	& 6 		\cr
MBM 36 		 & 2FGL J1553.5-0324$\;\;$ 	 & -2.22 & 15.85  &		---			&    ---		\cr
G313.1-28.6\tablenotemark{c} & 2FGL J1942.7-8049c  & -2.40  & 32.5  &		---			& 2, 6 \cr
G315.1-29.0\tablenotemark{c} & 2FGL J1925.7-7836c  & -2.31 & 14.1  &		---			& 1, 6 \cr
\enddata
\tablecomments{The molecular clouds coincident with point sources in the 2FGL. Properties are from the 2FGL. A blank means there is no classification or flag during the fitting process. Relevant flags are \citep{Fermicat2012}: \newline $N=1$: Source not detected significantly when the diffuse model was changed.\newline $N=2$: Source location changed beyond its 95\% error ellipse when the diffuse model was changed. \newline $N=6$: On top of an interstellar gas clump in the model of diffuse emission.}
\tablenotetext{a}{A source with a variability index above 41.6 is considered variable at the 99\% confidence level \citep{Fermicat2012}.}
\tablenotetext{b}{\citet{Fermicat2012} designation: bzb, tentative blazar classification. Coincident with the radio source MG3 J025805+2029}
\tablenotetext{c}{Part of Cham-East {\small II}}
\label{tab:ptsrces}
\end{deluxetable}

\clearpage

\begin{deluxetable}{l c c c c c c}
\tablewidth{0pt}
\tablecaption{Model Test Statistic Values}
\tablehead{
ROI name		&
$TS_\htwo$		&
$TS_\mathrm{DG}	$	&
$TS_\mathrm{CO}$	&
$TS_\mathrm{ex}$	&
$TS_\mathrm{AGN}$		
}
\startdata 
MBM 36			& 595 & 140 & 92 & 314 & 9\cr
DIR 071-43		& 76 & 47 & 2 & 76 & 7\cr
MBM 55			& 341& 0 & 273 & 318 & -2\cr
MBM 02			& 62 & 10 & 17 & 41 & 1\cr
MBM 04			& 80 & 47 & 18 & 80 & 18\cr
MBM 32			& 166 & 9 & 76 & 122 & 6\cr
MBM 12			& 691 & 37 & 346 & 308 & 96\cr
MBM 20			& 457 & 43 & 63 & 124 & 0\cr
Cham-East {\small II}& 832 & 93 & 453 & 624 & 16\cr
Control\tablenotemark{a}& 1 & 0 & -1 & -- & -- \cr
\enddata
\tablecomments{The significance of various diffuse emission components, rounded down to the nearest integer.}
\tablenotetext{a}{In the control region, the models with point sources were not fit because there was no significant CO emission and little dark gas in the ROI.}
\label{tab:TSvals}
\end{deluxetable}

\clearpage

\begin{deluxetable}{lcccc}
\tablewidth{0pt}
\tablecaption{Cloud Parameters}
\tablehead{
ROI Name 			&
$q_{\hi}$ 			&
$q_\mathrm{CO}$ 	&
$q_{\Avres}$ 			&
Photon Index }
\startdata 
MBM 36 			& $9.1\pm 0.5 ^{+2.0}_{-1.5}$	& $1.2\pm 0.1\pm0.2$  & 		$3.2\pm 0.3\pm0.5$	& $-2.13\pm0.10$\cr
DIR 071-43 		& 5.9$\pm0.3^{+1.3}_{-1.0}$ 	& $1.1\pm0.7\pm0.2$ 	& $1.6\pm0.3\pm0.3$ 	& $-2.17\pm 0.44$\cr
MBM 55 			& $8.9\pm0.9^{+2.0}_{-1.5}$ 	& $1.4\pm0.1\pm0.2$	& $0.3\pm0.2\pm0.1$ 	& $-2.23\pm 0.06$ \cr
MBM 02 			& $6.9\pm0.6^{+1.6}_{-1.2}$ 	& $2.4\pm0.7\pm0.4$	& $0.8\pm0.3\pm0.2$	& $-2.40\pm 0.19$\cr
MBM 04 			& $7.9\pm0.5^{+1.8}_{-1.3}$	& --- 				& $1.9\pm0.4\pm0.3$	& $-2.34\pm 0.11$\tablenotemark{a}\cr
MBM 32 			& $7.1\pm0.4^{+1.6}_{-1.2}$ 	& $2.1\pm0.2\pm0.3$	& $0.5\pm0.2\pm0.1$	& $-2.27\pm 0.07$ \cr
MBM 12 			& $12.4\pm0.2^{+2.8}_{-2.1}$	& $1.4\pm0.1\pm0.2$ 	& $1.3\pm0.2\pm0.2$	& $-2.05\pm 0.08$ \cr
MBM 20 			& $7.7\pm0.4^{+1.7}_{-1.3}$ 	& $2.8\pm0.4\pm0.4$	& $1.2\pm0.2\pm0.2$	& $-2.47\pm 0.10$ \cr
Cham-East {\small II}& $6.9\pm0.3^{+1.6}_{-1.2}$ 	& $1.3\pm0.1\pm0.2$ 	& $4.3\pm0.5\pm0.4$ 	& $-2.32\pm 0.06$ \cr
Control			& $9.0\pm0.5^{+2.0}_{-1.5}$ & --- & --- & $-2.61\pm 0.05$\tablenotemark{b} \cr 
%Average\tablenotemark{c}& $8.1\pm0.2^{+0.5}_{-0.3}$ & $1.6\pm0.1\pm0.3$ & $1.5\pm0.1\pm0.1$ & $-2.32\pm0.14$
\enddata
\tablecomments{Gamma-ray emissivity of gas templates for 250 MeV $<$ E $<$ 10 GeV with the associated statistical and systematic uncertainties. The photon index reported is from the CO template.\\Units: $q_{\rm \hi}$($10^{-27}$ photons \pers\ sr$^{-1}$ H-atom$^{-1}$), $q_\mathrm{CO}(10^{-6}$ photons \cmtwo\ \pers\ sr$^{-1}$ (K \kms)$^{-1}$), $q_{\rm \Avres}(10^{-5}$ photons \cmtwo\ \pers\ sr$^{-1}$ mag$^{-1}$).}
\tablenotetext{a}{CO not detected significantly, index taken from dark gas.}
\tablenotetext{b}{Neither CO nor dark gas detected significantly, index taken from \hi\ above $\sim$1 GeV.}
%\tablenotetext{c}{Average does not include control region.}
\label{tab:fit}
\end{deluxetable}

\clearpage

\begin{deluxetable}{lccc}
\tablewidth{0pt}
\tablecaption{Other Fitted Parameters}
\tablehead{
ROI Name 			&
$q_{\mathrm{H{\small I}}\_\mathrm{far}}$ 			&
$c_\mathrm{IC}$ 	&
$c_\mathrm{iso}$  }
\startdata 
MBM 36 			& --- & $1.2\pm0.7$ & $0.7\pm0.1$ \cr
DIR 071-43 		& --- & $2.6\pm0.3$ & $0.9\pm0.1$ \cr
MBM 55 			&  $11.2\pm1.8$ & --- & $1.3\pm0.1$ \cr
MBM 02 			& --- & $2.0\pm0.6$ & $1.1\pm0.1$ \cr
MBM 04 			& --- & $2.0\pm0.5$ & $1.0\pm0.1$ \cr
MBM 32 			& $12.4\pm1.2$ & $1.2\pm0.2$ & $0.8\pm0.1$ \cr
MBM 12 			& --- & --- & $0.7\pm0.1$ \cr
MBM 20 			& --- & $1.5\pm0.4$ & $1.0\pm0.1$ \cr
Cham-East {\small II}& --- & --- & $1.3\pm0.1$\cr
Control			& $13.6\pm1.6$ & --- & $1.0\pm0.1$ \cr 
\enddata
\tablecomments{Gamma-ray emissivities for the non-local \hi\ and the normalization factors for the inverse Compton and isotropic components for 250 MeV $<$ E $<$ 10 GeV with the associated statistical uncertainties. Blanks indicate the component was not detected significantly and thus not included in the model. \\Units: $q_{\mathrm{H{\small I}}\_\mathrm{far}}$($10^{-27}$ photons \pers sr$^{-1}$ H-atom$^{-1}$)}
\label{tab:fit_ICiso}
\end{deluxetable}

%%%% Kill units because they are in the text!

\clearpage

\begin{deluxetable}{lcccc}
\tablewidth{0pt}
\tablecaption{MBM 12 Parameters}
\tablehead{
Energies			&
$q_\mathrm{\hi}$ 			&
$q_\mathrm{CO}$ 		&
$q_\mathrm{\Avres}$ 			&
$c_\mathrm{iso}$  \cr
(MeV) & &&&}
\startdata 
250 -- 400 	& $3.24\pm0.44^{+0.74}_{-0.55}$ 	& $0.24\pm0.09\pm0.04$ 	& $1.46\pm0.26\pm0.25$ 	& $1.4\pm0.2$\cr
400 -- 630	& $2.45\pm0.19^{+0.55}_{-0.42}$	& $0.30\pm0.05\pm0.05$	& $0.45\pm0.14\pm0.08$	& $1.3\pm0.2$\cr
630 -- 1000	& $1.54\pm0.15^{+0.35}_{-0.26}$	& $0.21\pm0.03\pm0.03$	& $0.33\pm0.08\pm0.06$	& $1.3\pm0.3$\cr
1000 -- 1580	& $0.88\pm0.04^{+0.20}_{-0.15}$	& $0.08\pm0.02\pm0.01$	& $0.22\pm0.05\pm0.04$	& $1.0\pm0.1$\cr
1580 -- 2510	& $0.52\pm0.07^{+0.12}_{-0.09}$	& $0.07\pm0.01\pm0.01$	& $0.09\pm0.03\pm0.02$	& --- \cr
2510 -- 3980	& $0.19\pm0.03^{+0.04}_{-0.03}$	& $0.04\pm0.01\pm0.006$	& $0.02\pm0.01\pm0.003$	& $1.4\pm0.4$\cr
3980 -- 10000 & $0.08\pm0.03^{+0.02}_{-0.01}$	& $0.03\pm0.01\pm0.005$	& $0.03\pm0.01\pm0.005$	& $1.8\pm0.5$
\enddata
\tablecomments{Gamma-ray emissivities and the associated statistical and systematic uncertainties for the gas templates. The isotropic component only includes statistical uncertainties. Non-local \hi\ and inverse Compton are not detected significantly in any individual energy bin and are not included.}
\label{tab:fit_energy}
\end{deluxetable}

\clearpage
\begin{deluxetable}{lccccc}
\tablewidth{0pt}
\tablecaption{Calculated Cloud Properties}
\tablehead{
ROI Name 					&
\xco $/10^{19}$				&
$\xav/10^{20}$ 	&
$X_{\rm CO} /10^{20}$ 		\\
 							&
(\cmtwo\ (K \kms)$^{-1}$)		&
(\cmtwo\ mag$^{-1}$)			&
(\cmtwo\ (K \kms)$^{-1}$)		}
\startdata 
MBM 36 			& $6.8\pm 2.0$	 & $17.6\pm4.4$ 	& $1.4\pm 0.5 $ \cr
DIR 071-43 		& $9.3\pm6.7$ 	 & $13.6\pm4.0$ 	& $1.9\pm 0.9$ \cr
MBM 55 			& $7.6\pm2.3$  	& $1.7\pm1.2$ 	& $0.8\pm 0.2$ \cr
MBM 02 			& $17.5\pm7.0$ 	& $5.8\pm2.6$	& $2.3\pm 0.8$ \cr
MBM 04\tablenotemark{a}   	& ---    	& $11.8\pm3.8$ 	& --- \cr
MBM 32 			& $14.8\pm4.5$ 	& $3.5\pm1.6$ 	& $1.5\pm 0.4$ \cr
MBM 12 			& $5.5\pm1.8$   	& $5.4\pm1.5$ 	& $0.7\pm 0.2$ \cr
MBM 20 			& $18.0\pm5.6$  	& $7.8\pm2.2$ 	& $2.5\pm 0.8$ \cr
Cham-East {\small II}& $1.1\pm2.6$  	& $35.3\pm9.1$ 	& $1.7\pm 0.7$ \cr
\enddata
\tablecomments{Values for the conversion factors between \WCO\ and dark gas. The last column is the average value across the cloud of the combination of \xco\ and \xav\ as in Equation \ref{eq:x-def} assuming $f=1$. Uncertainties include systematics from the emissivities.}
\tablenotetext{a}{CO not detected significantly.}
\label{tab:calc}
\end{deluxetable}

\end{document}